\documentclass[aoas]{imsart}

\RequirePackage{amsthm,amsmath,amsfonts,amssymb}
\RequirePackage[authoryear]{natbib}
\RequirePackage[colorlinks,citecolor=blue,urlcolor=blue]{hyperref}
\RequirePackage{graphicx}
\RequirePackage{float}
\RequirePackage{algorithm}
\RequirePackage{algpseudocode}
\RequirePackage{underscore}
\RequirePackage{listings}
\RequirePackage{xcolor}
\usepackage{pdfpages}
\startlocaldefs
\theoremstyle{plain}

\theoremstyle{remark}

\definecolor{codegreen}{rgb}{0,0.6,0}
\definecolor{codegray}{rgb}{0.5,0.5,0.5}
\definecolor{codepurple}{rgb}{0.58,0,0.82}
\definecolor{backcolour}{rgb}{0.95,0.95,0.92}

\lstdefinestyle{mystyle}{
    backgroundcolor=\color{backcolour},   
    commentstyle=\color{codegreen},
    keywordstyle=\color{magenta},
    numberstyle=\tiny\color{codegray},
    stringstyle=\color{codepurple},
    basicstyle=\ttfamily\scriptsize,
    breakatwhitespace=false,         
    breaklines=true,                 
    captionpos=b,                    
    keepspaces=true,                 
    numbers=left,                    
    numbersep=5pt,                  
    showspaces=false,                
    showstringspaces=false,
    showtabs=false,                  
    tabsize=2
}

\endlocaldefs

\begin{document}

\begin{frontmatter}
\title{Synthesizing data products, mathematical models, and observations for lake temperature forecasting}
\runtitle{Lake temperature forecasting}

\begin{aug}
\author[A]{\fnms{Maike F. Holthuijzen}\ead[label=e1]{mholthuijzen@alumni.uidaho.edu}\orcid{0000-0002-6870-3314} },
\author[B]{\fnms{Robert B. Gramacy}\ead[label=e2]{rbg@vt.edu}}
\author[D]{\fnms{Cayelan C. Carey}\ead[label=e3]{cayelan@vt.edu}}
\author[B]{\fnms{David M. Higdon}\ead[label=e4]{dhigdon@vt.edu}}
\and
\author[C]{\fnms{R. Quinn Thomas}\ead[label=e5]{rqthomas@vt.edu}}

\address[A]{Virginia Tech \& Sandia National Labs Dept 8738\printead[presep={,\ }]{e1}}
\address[B]{Department of Statistics,
Virginia Tech\printead[presep={,\ }]{e2}\printead[presep={,\ }]{e4}}
\address[D]{Department of Biological Sciences,
Virginia Tech\printead[presep={,\ }]{e3}}
\address[C]{Departments of Forest Resources \& Environmental Conservation and Biological Sciences,
Virginia Tech\printead[presep={,\ }]{e5}}
\end{aug}

\begin{abstract}
We present a novel forecasting framework for lake water temperature, which is crucial for managing lake ecosystems and drinking water resources. The General Lake Model (GLM) has been previously used for this purpose, but, similar to many process-based simulation models, it: requires a large number of inputs, many of which are stochastic; presents challenges for uncertainty quantification (UQ); and can exhibit model bias. To address these issues, we propose a Gaussian process (GP) surrogate-based forecasting approach that efficiently handles large, high-dimensional data and accounts for input-dependent variability and systematic GLM bias. We validate the proposed approach and compare it with other forecasting methods, including a climatological model and raw GLM simulations. Our results demonstrate that our bias-corrected GP surrogate (GPBC) can outperform competing approaches in terms of forecast accuracy and UQ up to two weeks into the future.
\end{abstract}


\end{frontmatter}

\section{Introduction}
\label{sec:intro}

\begin{figure}[ht!]
\centering
\begin{minipage}{4.7cm}
\includegraphics[scale=0.60,trim=200 0 50 15]{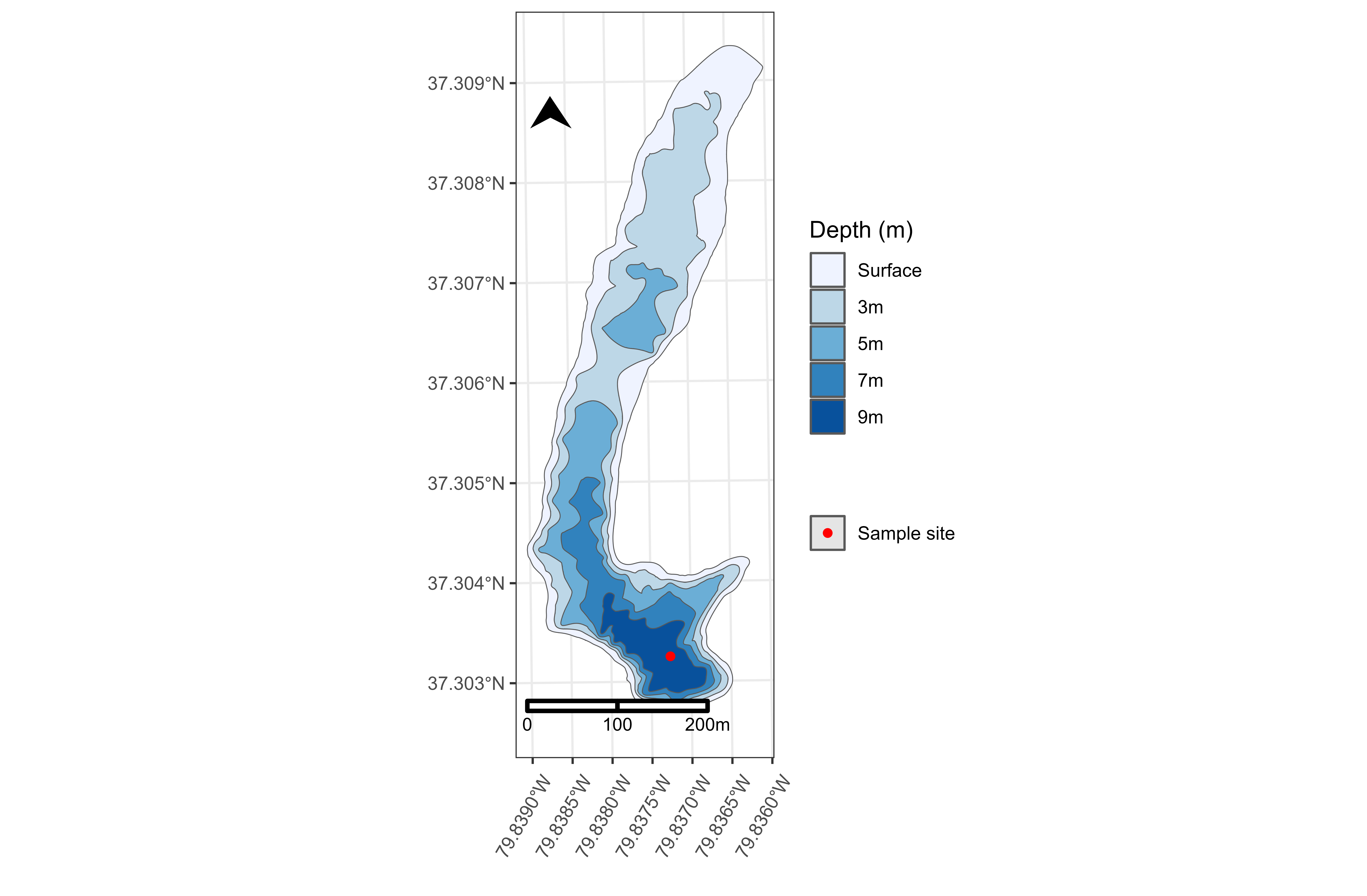} 
\end{minipage}
\hspace{0.7cm}
\begin{minipage}{8.5cm}
\includegraphics[scale=0.43]{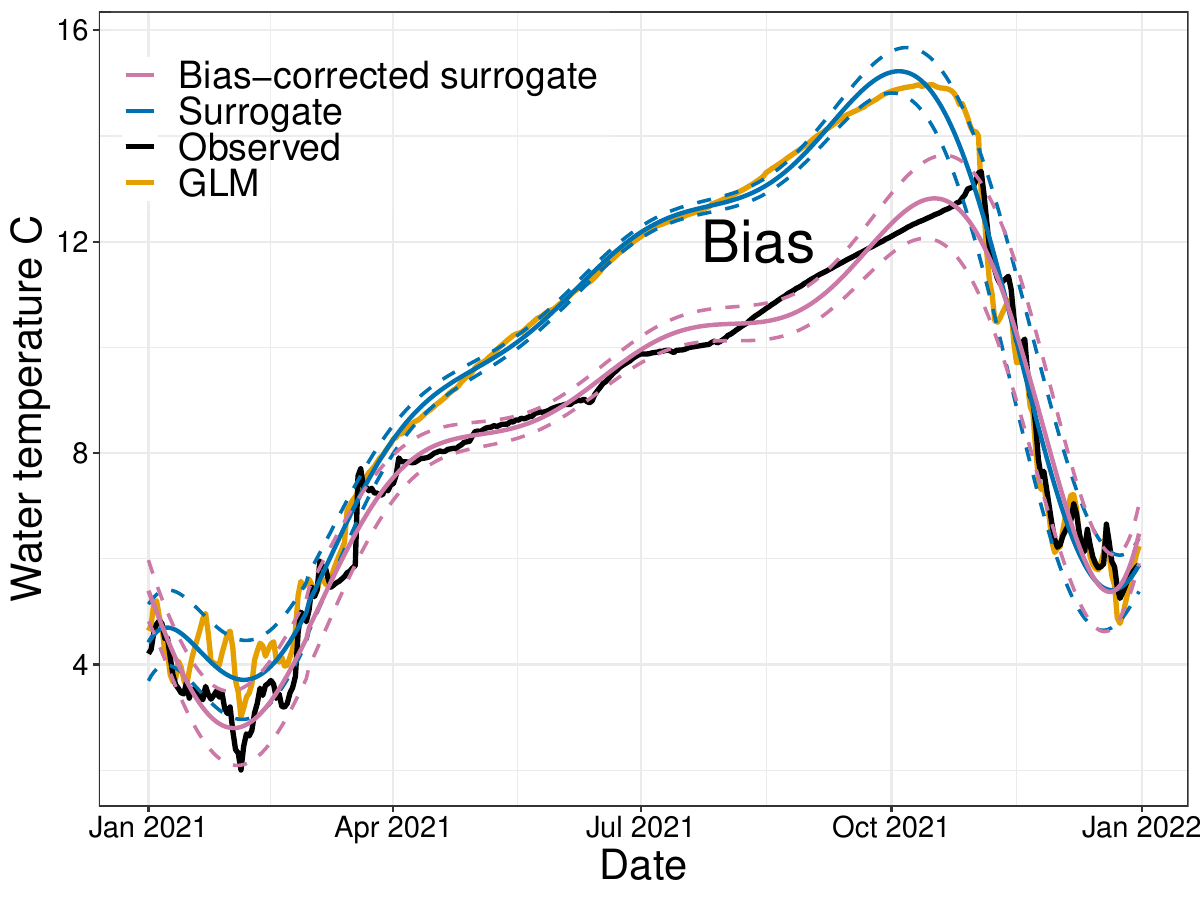} 
\end{minipage}
\vspace{-0.5cm}
\caption{{\em Left:} Falling Creek Reservoir (FCR; 37.30$^{\circ}$N, 79.84$^{\circ}$W). 
{\em Right:} Example GLM simulations (lighter thick lines) and observations (thick dark lines) for 2021 at 6m; means and 90\% PIs of a GLM surrogate (thin lines); a bias-corrected (light, thin lines).} \label{fig:PlotForCayelanannotate}\label{fig:FCRplot}
\end{figure}

Process-based models are critical for understanding, managing, and forecasting environmental phenomena and are widely used in ecological research. 
They are crucial for forecasting lake water temperature profiles (e.g., at varying depth) that influence decisions that impact the health of aquatic ecosystems \citep{carey2022advancing}.  Water quality and the health of aquatic organisms are strongly associated with water temperature \citep{carey2022advancing}, which influences the growth and distribution of aquatic flora and fauna \citep{wetzel2001limnology}. Warmer water can accelerate the growth of harmful phytoplankton, threatening aquatic and human health \citep{carey2012eco}. Lake temperature forecasts that extend at least 30 days into the future, include uncertainty quantification (UQ), and are updated frequently are crucial to adequately managing drinking water resources, mitigating water quality degradation, and planning for the health of lake ecosystems \citep{thomas2023near,carey2022advancing}. Here, we focus on forecasting lake temperatures at ten depths at Falling Creek Reservoir (FCR), a small reservoir in Vinton, Virginia, USA (Figure \ref{fig:FCRplot} \emph{left}). 

There are several approaches to forecasting lake temperatures \citep{lofton2023progress}, the simplest of which are climatological. These are based on historical data, capturing broader trends, and are accurate at long (15+ day) horizons \citep{thomas2023near}. Such forecasts for FCR can be constructed using data collected by \textit{in situ} water temperature sensors (denoted by the red dot in Figure \ref{fig:FCRplot} \emph{left}). Climatological forecasts do not vary substantively year upon year, their predictive limits do not vary with the forecast horizon, and they lack sharpness 
reducing their value for near-term water quality management.

Lake ecosystem models are designed to simulate heat transfer within aquatic ecosystems 
and can encompass a wide range of processes, including hydrodynamics, 
water quality,
nutrient cycling,
and ecological interactions. 
One such model, the one-dimensional (1D) General Lake Model \citep[GLM;][]{hipsey2019general},\footnote{Apologies for the acronym; here ``GLM'' refers to the General Lake Model, not a generalized linear model.} can simulate thermal dynamics in a vertical column of a waterbody. GLM takes into account the physical characteristics of the lake, such as its depth, surface area, and geometry, water inputs from surrounding areas, and external factors (e.g., meteorology) to simulate water movement, mixing, and circulation patterns within the water column \citep{hipsey2019general}. GLM has been used in many contexts, which we do not review here.
However, the GLM has characteristics that make it challenging to use for near-term (1--30 days) forecasts of lake temperatures. Since GLM is a deterministic model, UQ associated with forecasts can only be derived if it is driven with an \emph{ensemble} of model inputs, such as weather forecasts or model parameters. Importantly, like all process-based models, 
GLM is a simplified version of reality that does not perfectly represent processes that govern energy transfer within a water column. 
Any systematic biases, or other errors that result from such oversimplifications, need to be accounted for.

Observations provide a critical foundation for assessing the accuracy of GLM. At FCR, water temperature sensors installed at 10 depths (surface to 9m depth by 1m increments) provide a rich and continuous observational data record starting in July 2018 (dot in Figure \ref{fig:PlotForCayelanannotate} {\em left}). One can get a sense of GLM performance for the FCR study region by comparing GLM output to sensor data. Figure \ref{fig:PlotForCayelanannotate} ({\em right)} shows an example of GLM simulations (thick, light line) and sensor observations (thick, dark line) at 6m depth 
as well as a \emph{surrogate model} for GLM in thinner  lines (discussed momentarily). These GLM simulations exhibit a pronounced warm bias between April and October with respect to observations. 

\subsection*{Our contributions}

\begin{figure}[ht!]
\centering
\includegraphics[scale=0.47, trim=0cm 1.5cm 0cm 0cm]{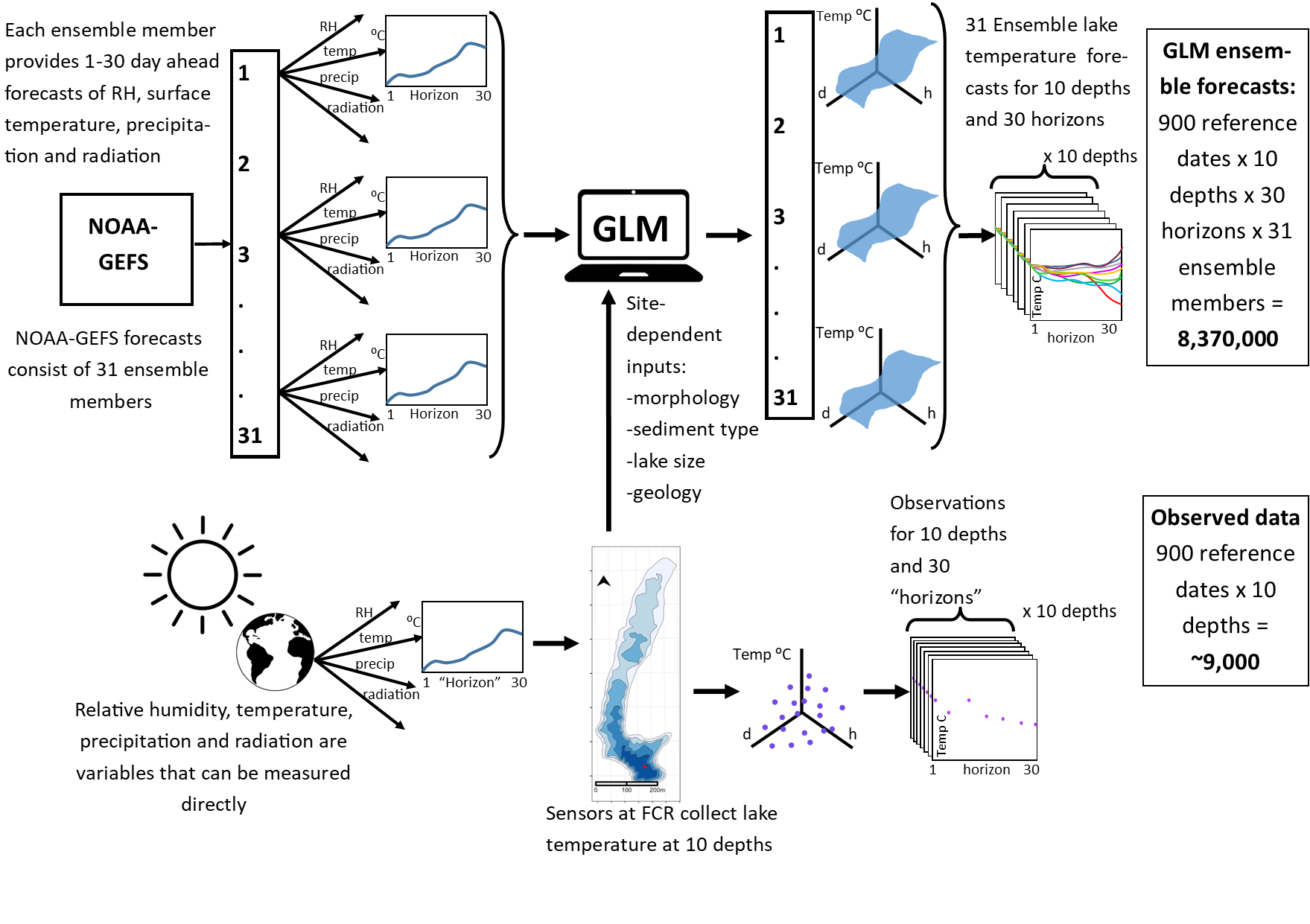}
\caption{Schematic relating NOAA forecasts to GLM simulations to FCR sensor data sensor data towards deriving 1--30 day ahead forecasts at 10 depths from 2020-10-03 to 2023-06-10.} 
\label{fig:GiantSchematic2}
\end{figure}

We seek accurate 1-30 day ahead ensemble forecasts of lake temperatures at 10 depths at FCR with well-calibrated UQ. These forecasts are made by driving GLM with a 31-member ensemble of weather forecasts generated by the National Oceanic and Atmospheric Administration Global Ensemble Forecast System \citep[NOAA-GEFS;][see \url{https://www.ncei.noaa.gov/products/weather-climate-models/global-ensemble-forecast}]{hamill2022reanalysis}. Construction of ensemble lake temperature forecasts is facilitated with automated, publicly available code (\url{https://github.com/maikeh7/Surrogate_Assisted_GLM}) and is inspired by \citet{thomas2020near}. Figure \ref{fig:GiantSchematic2} diagrams how NOAA ensemble forecasts and data from FCR are used to construct lake temperature forecasts with GLM.

\begin{figure}[ht!]
\centering
\includegraphics[scale=0.62]{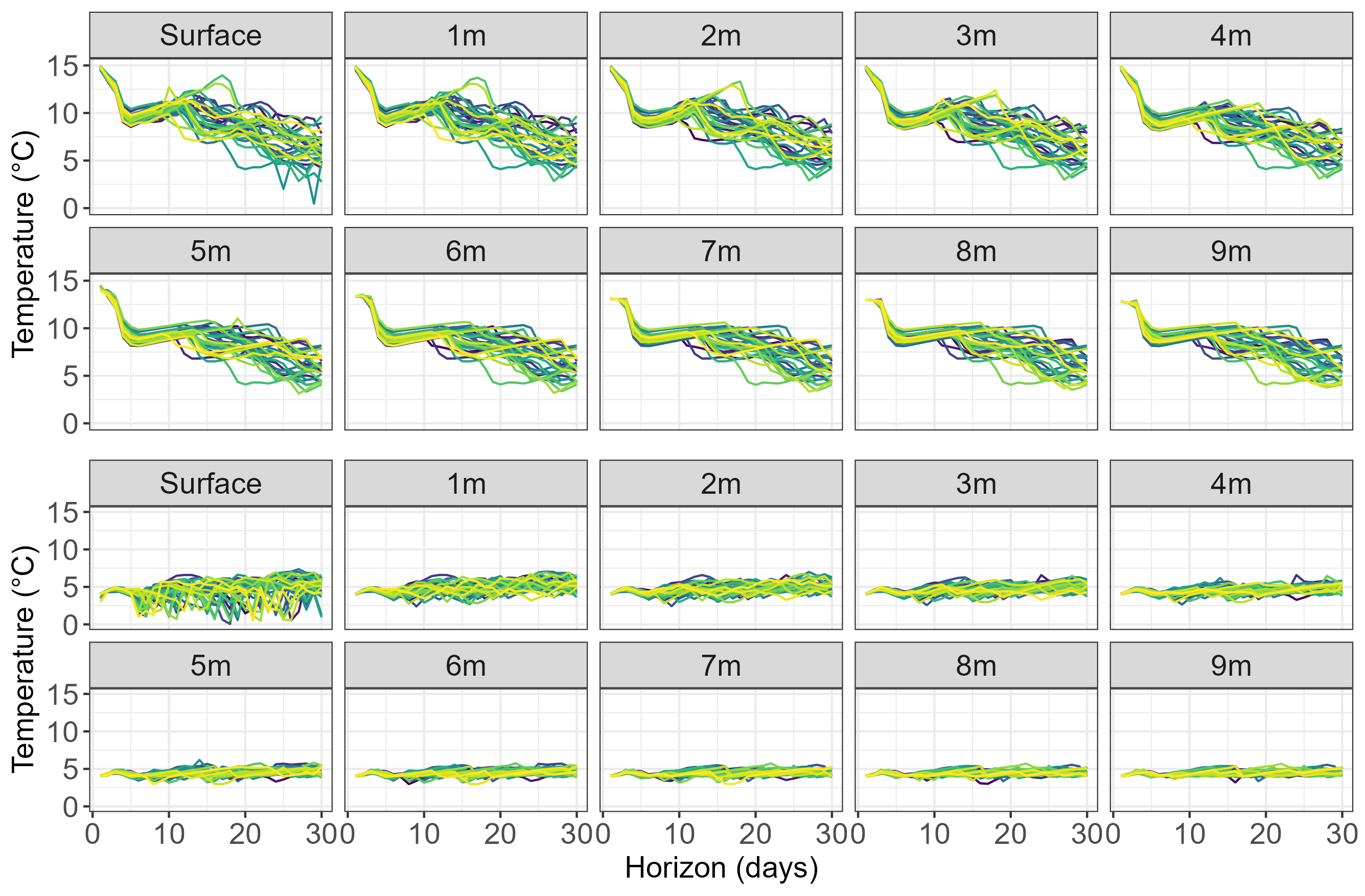}
\vspace{-0.25cm}
\caption{NOAA-GLM lake temperature forecasts ($^{\circ}$C) for FCR starting on reference dates A) 2020-10-30 and B) 2021-01-01. Different shades denote trajectories of ensemble members. Panel numbers denote depth (m).}
\label{fig:NOAA1}
\end{figure}

Examples of NOAA-driven GLM forecasts (hereafter ``NOAA-GLM'') at FCR for 1--30 days in the future are shown in Figure \ref{fig:NOAA1}. At lower horizons, the forecasts of individual ensemble members are similar, but as horizon increases, forecasts diverge. There is considerable variability within reference dates: forecasts starting on 2020-10-30 (fall, Figure \ref{fig:NOAA1}A) are more variable compared to forecasts originating on 2021-01-01 (winter, Figure \ref{fig:NOAA1}B). Variability also differs according to lake depth, where shallow depths exhibit more variation than deeper ones. While simple, this approach is useful: forecasts have long (30-day) horizons, and uncertainty can be derived from variability among ensemble members. However, it does not account for GLM bias, and uncertainty is not modeled directly, making UQ less reliable. 

A more robust alternative to generating lake temperature forecasts involves building a \emph{probabilistic} representation of NOAA-GLM, which could subsequently improve prediction accuracy and UQ, while also denoising the 31 ensemble forecast trajectories. This approach could accommodate functional forms of bias and correction. Modeling bias is also advantageous because it allows the bias to be separated from observational noise. 
Observe in Figure \ref{fig:PlotForCayelanannotate} how one type of statistical model (the surrogate) adequately smooths GLM simulations while providing appropriate UQ. We propose to extend this idea from single to ensemble GLM forecast trajectories. Figure \ref{fig:PlotForCayelanannotate} ({\em right}) also shows the mean and 90\% prediction intervals resulting from a \emph{bias-corrected} model. In contrast to the biased surrogate model fitted to GLM simulations, the mean of the bias-corrected model provides a very good approximation to observations, and UQ is reliable.

One way to statistically model NOAA-GLM forecasts is through the use of a computer model \emph{surrogate}. A surrogate emulates the \emph{behavior} of a computer model such as GLM but does so without the model's physical equations \citep{santner2018design}. 
We propose using a surrogate to model NOAA-GLM lake temperature forecasts. 
Gaussian process (GP) surrogates are the canonical choice as they have many desirable properties, including flexibility, accuracy, and reliable UQ \citep{gramacy2020surrogates, williams2006gaussian}. Bias can also be modeled with a GP \citep{kennedy2001bayesian}. One major challenge to using a GP surrogate is the decomposition bottleneck of multivariate normal (MVN) covariance matrices.  In GP fitting/prediction, these matrices are sized commensurately with the square of the training data of size $n$, incurring a decomposition cost that is cubic in $n$ flops. In
our application, $n$ is very large.  Each of the 31 NOAA ensemble members results in 31 NOAA-GLM forecasts for 30 horizons and 10 lake depths, producing a total dataset size of: 900 reference dates $\times$ 30 horizons $\times$ 10 depths $\times$ 31 ensemble members $\rightarrow n= 8{,}370{,}000$, a formidable data size for any model-fitting enterprise, especially for GP surrogates. 


GP surrogates have been used in hydrological \citep{razavi2012review} and environmental applications \citep{fer2018linking}, model calibration \citep{fer2018linking}, sensitivity  \citep{oakley2004probabilistic} and uncertainty analysis \citep{yang2018uncertainty, roy2018comparison}. However, \emph{bias-corrected}, GP surrogates that can handle input-dependent variability and big data have never been implemented in the context of ecological forecasting.  We aim to generate bias-corrected forecasts (10 depths and 30 horizons) via NOAA-GLM ensembles fast enough so updates for all quantities may be furnished daily. The remainder of the paper is organized as follows: we begin with a review of GP surrogates for computer models and follow with a detailed description of the forecasting workflow and the statistical techniques associated with it. We validate our proposed framework and show how forecasts from our framework compare to one in which bias is not corrected, raw GLM output, and a climatological forecast. We then discuss the results and provide concluding remarks.

\section{Background and proof-of-concept}
\label{sec:back}

Our forecasting framework operates on the frontier of Gaussian process (GP) surrogate modeling \citep[e.g.,][]{gramacy2020surrogates} for synthesizing computer model output, observations, and correcting bias.  A basic GP setup is introduced here, building upon Figure \ref{fig:PlotForCayelanannotate}. This preliminary analysis suggests that forecasts utilizing GLM and lake temperatures measured by sensors would benefit from meta-modeling and bias correction. Ultimately, we will need a higher-powered strategy in order to cope with a large corpus of NOAA-GLM runs, which is the subject of
Section \ref{sec:method}.

\subsection{Gaussian process surrogate modeling}

GPs are popular as surrogates for computer model simulations  \citep[e.g.,][]{santner2018design}, and more widely for nonlinear regression \citep[e.g.,][]{williams2006gaussian}, because they furnish accurate predictors with appropriate coverage when used to model smooth
and stationary (i.e., physical) phenomena. Here, we utilize GPs in both contexts: for computer model simulations and for observational data.  

Consider a computer model $M$ represented as a function $f: \mathbb{R}^p \rightarrow \mathbb{R}$ that maps inputs to outputs, whose pairs over $n_M$ simulations comprise data $D_{n_M} = (\mathbf{x}_i, y_i)$, $i = 1 \dots n_M$.   
Let $\mathbf{X}_{n_M} = (\mathbf{x}_1^\top, \mathbf{x}_2^\top, \dots, \mathbf{x}_{n_M}^\top) \in \mathbb{R}^{n_M \times p}$, collect inputs,
and  $\mathbf{Y}^M_{n_M} = \left ( y_1, y_2, \dots, y_{n_M} \right )^\top$ outputs. 
Utilizing a GP model, or {\em prior}, for $f$ amounts to specifying that $\mathbf{Y}_{n_M}^M$ follow a multivariate normal distribution (MVN). It is common that the MVN be specified with a mean of zero, and a covariance $\Sigma$ that is determined
by (inverse) distances between inputs $\mathbf{X}_{n_M}$:
\begin{equation}\label{eq:Ydist}
\mathbf{Y}^M_{n_M} \sim \mathcal{N}_{n_M}(\mathbf{0}_{n_M}, \Sigma(\mathbf{X}_{n_M}))
\quad \mbox{where, e.g.,} \quad \Sigma^{ij} = \tau^2(k(q(\mathbf{x}_i, \mathbf{x}_j)) + g  \mathbb{I}_{\{i=j\}}).
\end{equation}
There are many choices for the form of $\Sigma(\mathbf{X}_{n_M})$. Most are similar to Eq.~(\ref{eq:Ydist}).
A {\em kernel} $k(\cdot)$ performs the inversion, like 
$k(q) = e^{-q}$, and $q(\cdot, \cdot)$ calculates a (scaled) Euclidean distance:
\begin{equation}\label{eq:qdist}
q(\mathbf{x}_i, \mathbf{x}_j) =  \left ( \sum_{\ell = 1}^p     
\frac{|| x_{i,\ell} - x_{j,\ell}||^2}{\gamma_{\ell}} \right )^{1/2}.
\end{equation}
Our setup is largely indifferent to variations in choices of $q$, $k$, etc. In our empirical work we use the $q$ specified in Eq.~(\ref{eq:qdist}), and the Mat\'ern kernel $k$ with smoothness level fixed at 3.5. 

In what follows we express the setup in Eq.~(\ref{eq:Ydist}) concisely as $\mathcal{GP} (D_{n_M})$, given a choice of $k(\cdot)$, distance $q(\cdot)$ and hyperparameters $\boldsymbol{\theta} = (\boldsymbol{\gamma}, g, \tau^2)$.  The $\boldsymbol{\gamma} = (\gamma_1, \gamma_2, \dots \gamma_{d})$ are so-called {\em lengthscale} or {\em range} parameters, where $1 / \gamma_{\ell}$ determines the ``relevance'' of the $x_{\ell}$ 
The overall amplitude of the response is determined by {\em scale} parameter $\tau^2$, and the {\em nugget} $g$ operates
as a dial partitioning that amplitude between signal $(\tau^2)$ and noise $(\tau^2 g)$.  
Observe that $g$ is only augmenting the diagonal of $\Sigma$, implementing an independent and constant (i.e., homoskedastic) noise component, an aspect that we relax in Section \ref{sec:method}.  

Ideally, settings for hyperparameters are learned from data. The GP model (\ref{eq:Ydist}) for $\mathbf{Y}^M_{n_M}$ gives rise to a likelihood through the density of an MVN:
\begin{equation}\label{eq:likelihood}
L(\boldsymbol{\theta} ; D_{n_M}) \propto |\Sigma_{\boldsymbol{\theta}}|^{-1/2} \mathrm{exp} \left( -\frac{1}{2} (\mathbf{Y}^M_{n_M})^\top \Sigma_{\boldsymbol{\theta}}^{-1} \mathbf{Y}^M_{n_M}   \right),
\end{equation}
which can be used to learn settings $\hat{\boldsymbol{\theta}}_{n_M}$; see, e.g., \citet[][Chapter 5]{gramacy2020surrogates}.
Observe that $\boldsymbol{\theta}$ is buried in $\Sigma_{\boldsymbol{\theta}}$, although we often drop the subscript to streamline notation. We presume  $\hat{\boldsymbol{\theta}}_{n_M}$ has been pre-estimated given data $D_{n_M}$ to focus on other aspects of inference. More detail on our own procedures for an expanded $\boldsymbol{\theta}$ is provided in Section 3. Note that determinants and inverses in Eq.~(\ref{eq:likelihood}) require cubic in $n_M$ flops for dense matrices $\Sigma_{\boldsymbol{\theta}}$.  

Suppose we wish to predict outputs at a new $n' \times d$ set of input configurations collected in the rows of $\mathcal{X}$. For computer simulations, predictive equations comprise the ``surrogate'', because these $\hat{f}(\cdot)$ may be used in lieu of a new simulation $f(\cdot)$.  In geostatistics these are known as the {\em kriging equations} \citep{banerjee2003hierarchical}.  
Deriving the distribution for $\mathbf{Y}(\mathcal{X})$ involves stacking the MVN for $\mathbf{Y}_{n_M}$ together with an analogous one for $Y(\mathcal{X})$, i.e., where the latter follows Eq.~(\ref{eq:Ydist}) but where $\Sigma(\mathcal{X})$ is used instead.  The joint, $(n_M + n')$-dimensional MVN has a covariance structure with block diagonal components $\Sigma(\mathbf{X}_{n_M})$ and $\Sigma(\mathcal{X})$, and off-diagonal $\Sigma(\mathbf{X}_{n_M}, \mathcal{X})$ and its transpose.  The quantity $\Sigma(\mathbf{X}_{n_M}, \mathcal{X})$ is derived from distances between training $\mathbf{X}_{n_M}$ and testing $\mathcal{X}$ but involves no nugget ($g$) augmentation.  Then,  MVN conditioning  provides $Y(\mathcal{X}) \mid D_{n_M} \sim 
 \mathcal{N}_{n'} (\mu_{n_M}(\mathcal{X}), \Sigma_{n_M} (\mathcal{X}))$, where
\begin{align}
\label{eq:GPpredeqns}
\mu_{n_M}(\mathcal{X}) &= k( \mathbf{X}_{n_M}, \mathcal{X})^\top (k(\mathbf{X}_{n_M}) + g \mathbb{I}_{n_M})^{-1} \mathbf{Y}_{n_M}, \\
\mbox{and } \quad \Sigma_{n_M} (\mathcal{X}) &= \hat{\tau}^2 ( k(\mathcal{X}) + g \mathbb{I}_{n'} - \nonumber
k(\mathbf{X}_{n_M}, \mathcal{X})^\top (k(\mathbf{X}_{n_M}) + g \mathbb{I}_{n_M})^{-1} k( \mathbf{X}_{n_M}, \mathcal{X})).
\end{align} 
Modulo $\hat{\boldsymbol{\theta}}$, these
equations provide full predictive uncertainty.  Any intervals derived from the quantiles of the covariance diagonal could be considered predictive intervals (PIs). 

If, instead, one is interested in predictions for $f(\mathcal{X}) = \mathbb{E}\{Y(\mathcal{X})\}$ one would take $g=0$ in $k(\mathcal{X}) + g \mathbb{I}_{n'}$ in Eq.~(\ref{eq:GPpredeqns}) to obtain confidence intervals (CIs) instead. More specifically,
and since we will need to use it later, if $\boldsymbol{\mu}_{n_M} \equiv \mu_{n_M}(\mathcal{X}) = \mathbb{E}\{Y(\mathcal{X}) \mid D_{n_M}\}$, then
\begin{equation}
 \mathbb{V}\mathrm{ar}\{ \boldsymbol{\mu}_{n_M} \} =
 \hat{\tau}^2 ( k(\mathcal{X}) - k(\mathbf{X}_{n_M},\mathcal{X})^\top (k(\mathbf{X}_{n_M}) + g \mathbb{I}_{n_M})^{-1} k( \mathbf{X}_{n_M}, \mathcal{X}))).  \label{eq:nonug}
\end{equation}
Note that $\mathbb{V}\mathrm{ar}\{ \boldsymbol{\mu}_{n_M} \} < \Sigma_{n_M} (\mathcal{X})$,
uniformly over all entries of those matrices.\footnote{Note also that a non-zero nugget $g$ is present elsewhere in Eq.~(\ref{eq:nonug}).}  Moreover $\mathbb{V}\mathrm{ar}\{ \boldsymbol{\mu}_{n_M} \} \rightarrow \boldsymbol{0}$ as $n_M \rightarrow \infty$ as long as $\mathbf{X}_{n_M}$ fills out the input space.  In other words, we can learn to emulate $f$ with the surrogate $\hat{f}$ perfectly with
enough data.  
We compare and contrast PIs and CIs in Figure \ref{fig:heteroex2}, alongside our heteroskedastic models in Section \ref{sec:sk}.   

\subsection{Gaussian process surrogates for lake temperatures}
\label{sec:GPlaketemps}
To illustrate, consider the following.
Take $f/M$ as GLM and run it for each day of 2021, driven by observed environmental variables in that year, and extract temperatures for ten depths: surface/0 to 9m.  Using the notation established above, $\mathbf{Y}_{n_M}$ is comprised of $n_M = 366 \times 10 = 3660$ simulation outputs, and $\mathbf{X}_{n_M}$ has $p=2$ columns indexing time $t$ in Julian day, and depth $d$, in meters, respectively.  Figure \ref{fig:PlotForCayelanannotate} (\emph{right})  shows a ``slice'' of {\em some} of the elements of this data ($D_{n_M}$) where $\mathbf{X}_{n_M}[\cdot,2] = 6$ meters.  (The full $D_{n_M}$ is ten times bigger.)

Now consider first modeling $D_{n_M} = (\mathbf{X}_{n_M}, \mathbf{Y}_{n_M})$ as a GP, i.e., $\mathcal{GP}(D_{n_M})$.  Specifically, estimate hyperparameters ($\boldsymbol{\theta} =( \boldsymbol{\gamma}, g, \tau^2)$) by MLE (\ref{eq:likelihood}), and plug these values $\hat{\boldsymbol{\theta}}$ into predictive equations (\ref{eq:GPpredeqns}) using $\mathcal{X} = \mathbf{X}_{n_M}$,
i.e., to evaluate the surrogate at the training locations. 
Such $n_M$, in the small tens of thousands, is on the cusp of manageable with modern architectures/libraries.
\begin{figure}[ht!]
\centering
\vspace{-0.5cm}
\includegraphics[scale=0.1,trim=75 20 20 0]{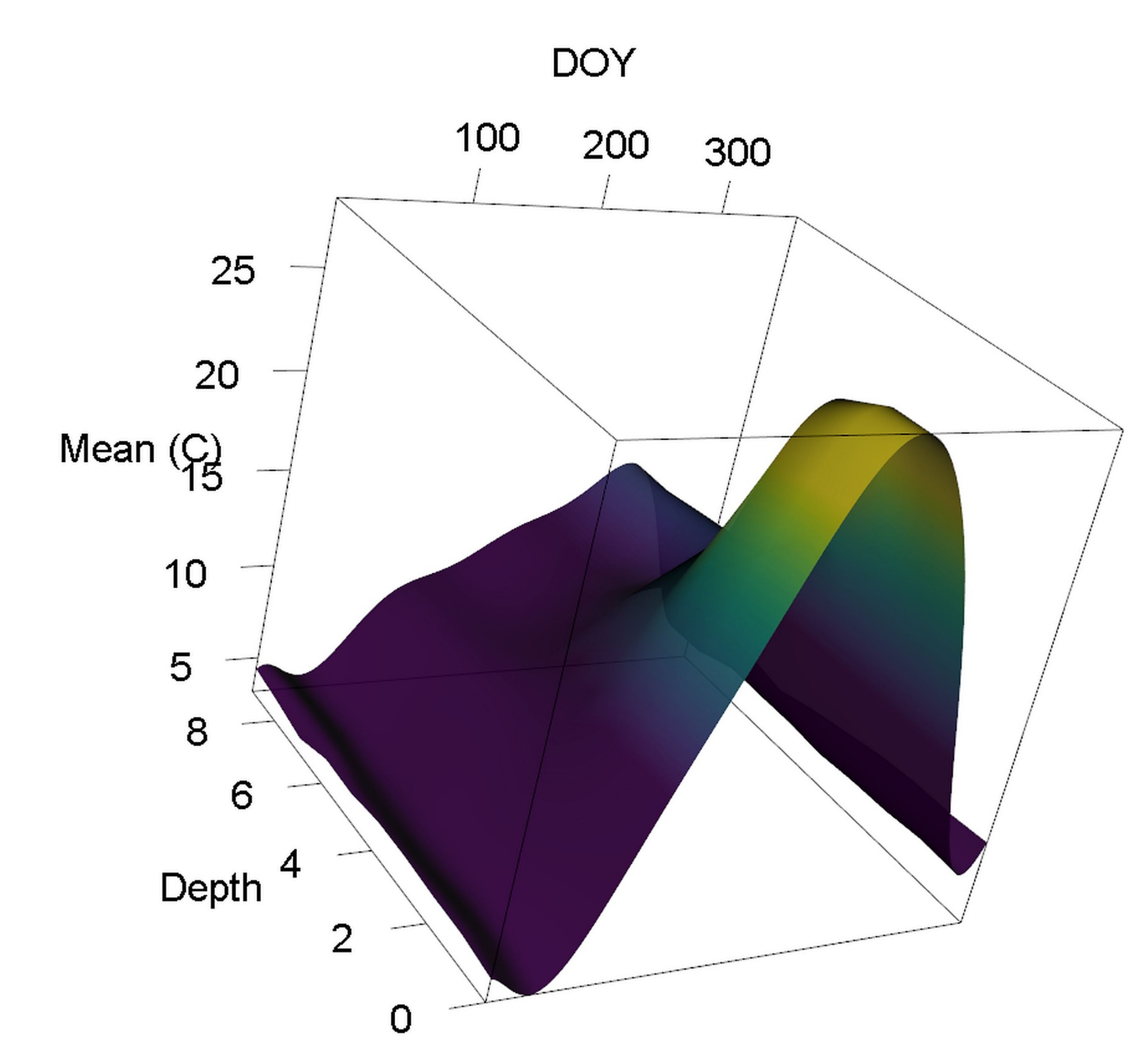} 
\includegraphics[scale=0.1, trim=70 20 20 0]{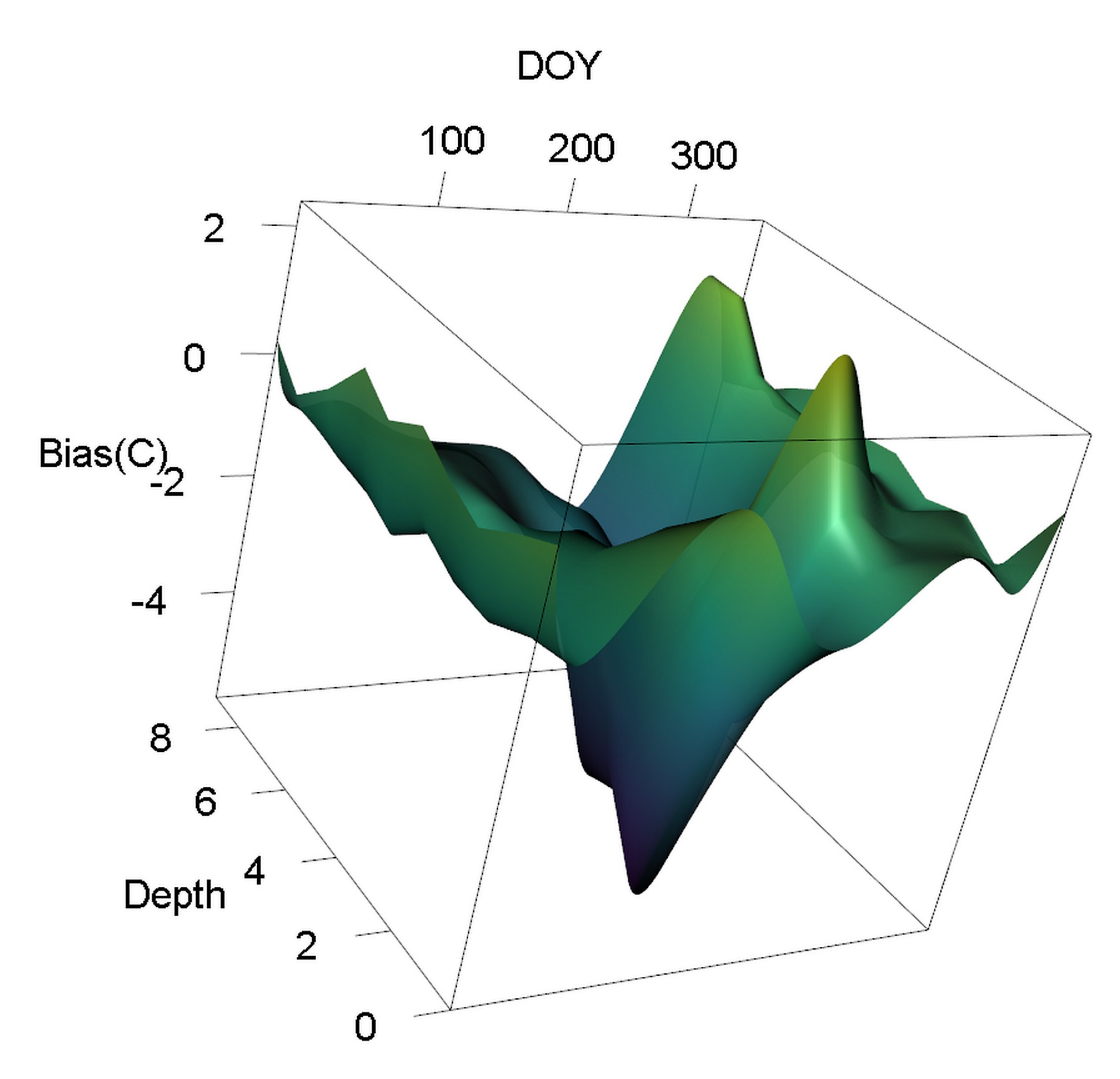} 
\includegraphics[scale=0.25, trim=0 0 0 0]{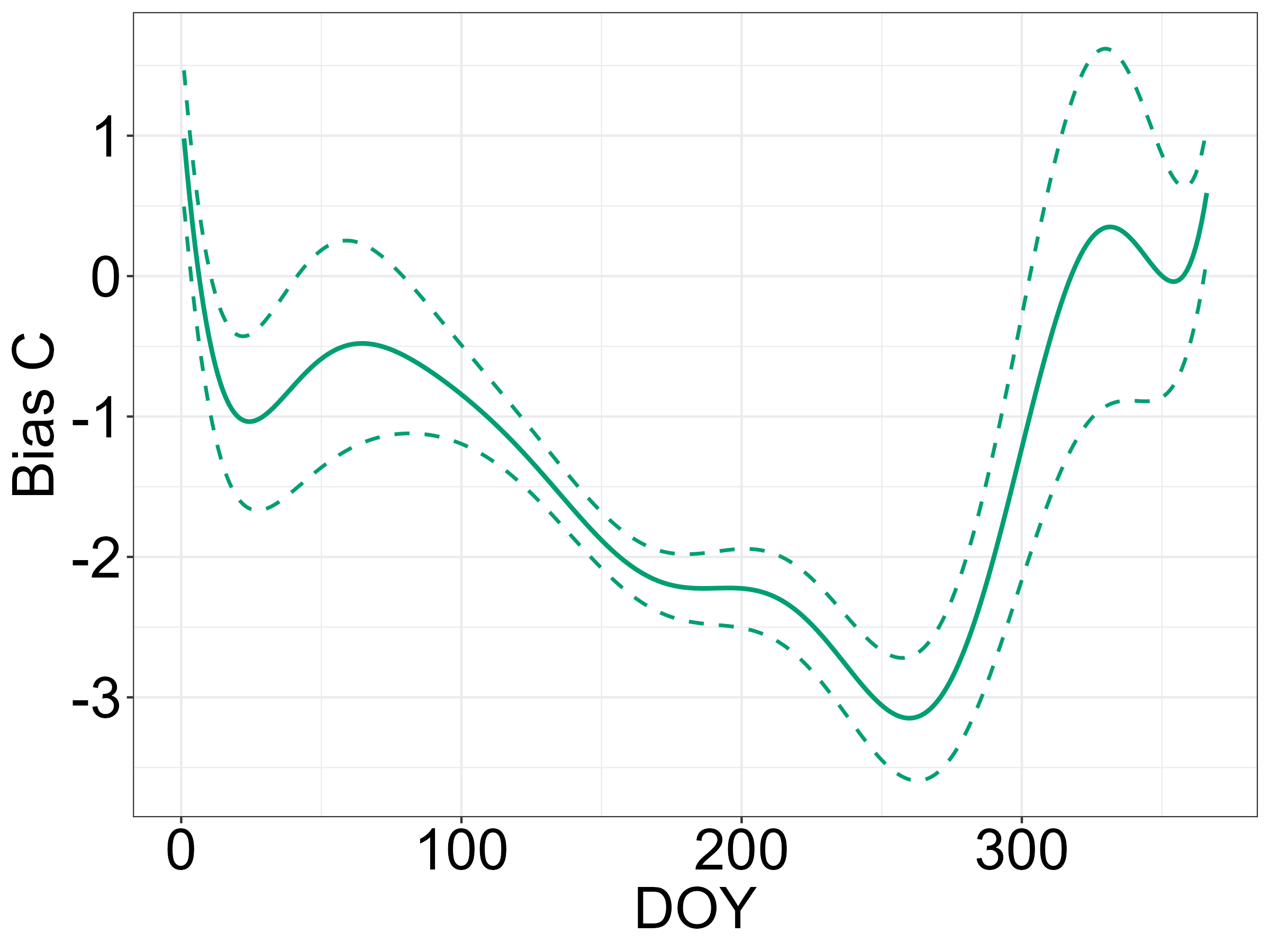} 
\vspace{-0.25cm}
\caption{ {\em Left:} GLM surrogate mean surface $\mu_{n_M}(\mathcal{X})$ over day of year (DOY) and depth (m) for 2021. {\em Middle:} mean surface $\boldsymbol{\mu}_{b}(\mathcal{X})$ of the bias GP $\hat{b}(\cdot)$. {\em Right:} 1D slice of $\boldsymbol{\mu}_{b}(\mathcal{X})$ via f $\mathcal{X}[\cdot,2] = 6$ meters, along with 90\% PIs. 
}\label{fig:MeanAndBias}
\end{figure}

Figure \ref{fig:MeanAndBias} ({\em left}) shows the resulting
surrogate mean $\mu_{n_M}(\mathcal{X})$ over depth (m) and day of year.  This corresponds to the dark lines in the right panel of Figure \ref{fig:PlotForCayelanannotate}, which provides a slice at 6m ($\mathcal{X}[\cdot,2] = 6$) along with 90\% PIs. Observe in Figure \ref{fig:PlotForCayelanannotate} (\emph{right}) how the surrogate smooths over GLM simulations and that intervals have appropriate coverage.  

Predictions from the surrogate $\hat{f}(\mathcal{X})$ in Figures \ref{fig:PlotForCayelanannotate} (\emph{right}) and  \ref{fig:MeanAndBias} (\emph{left}) are not ``forecasts'' because they do not correspond to the future. These predictions are derived from known or measured environmental variables, that, for forecasting, would be unknown or require additional modeling.  That is the subject of Section \ref{sec:method}. 


Importantly, $\hat{f}(\mathcal{X})$ in Figures \ref{fig:PlotForCayelanannotate}--\ref{fig:MeanAndBias} allow us to abstract raw GLM simulations by smoothing over noisy or chaotic measurements of environmental variables.  Like any statistical model, the surrogate helps us separate signal from noise.  Once we have the signal, we can compare it to something else, like actual physical/sensor measurements of lake temperature (black line in Figure \ref{fig:PlotForCayelanannotate}). From Figure \ref{fig:PlotForCayelanannotate} we see that GLM is biased from April to November.  In fact, narrow PIs here provide evidence of a strong, statistically significant bias
since the black lines go outside the blue-dashed PIs in early 2021. This warm bias would have been hard to separate from noise without the surrogate -- in particular without good UQ.  Meanwhile, in late 2021, the situation is different: GLM is likely unbiased; any differences with sensor measurements are probably due to noise.  

These are qualitative assessments that can be made concrete and quantitative by estimating bias as a function of time and depth, and using that estimate to make a correction (i.e., to obtain a ``bias-corrected surrogate'').  Let $\mathbf{Y}^F_{n_F}$ denote the $n_F$-length vector of sensor observations, or {\em field data}, at FCR.  In our example, $n_F = n_M$ since we have observed temperatures on the same days/depths as our GLM simulations ($\mathbf{X}_{n_F} = \mathbf{X}_{n_M}$).  We could easily have more simulations than sensor observations, say, and we will later in Section \ref{sec:method}.

Let the predictive mean of $\hat{f}(\mathcal{X})$ be denoted as $\boldsymbol{\mu}_{n_M} \equiv \mu_{n_M}(\mathcal{X})$, following Eq.~(\ref{eq:GPpredeqns}) with $\mathcal{X} = \mathbf{X}_{n_F}$. 
Using this quantity, we may measure the bias of our GLM surrogate 
as
\begin{equation}\label{eq:Calcbias}
    \mathbf{Y}^{b}_{n_F} = \mathbf{Y}^F_{n_F} - \boldsymbol{\mu}_{n_M}.
\end{equation}
We may then form a data set of observed discrepancies between model and field data  $D^b_{n_F} = (\mathbf{X}_{n_F}, \mathbf{Y}^b_{n_F})$.  Then, by regarding bias, or bias correction, $b(\cdot)$ as an unknown function like $f(\cdot)$, we may model it as $\mathcal{GP}(D^b_{n_F})$ and infer its hyperparameters, etc.  Readers familiar with the computer modeling literature will recognize a similarity between this setup and the so-called Kennedy and O'Hagan framework \citep[KOH;][]{kennedy2001bayesian} where the two processes: computer model $M$ and field data $F$ are meta-modeled jointly, without directly calculating discrepancies.  Our setup here is more modular, 
and thus more amenable to up-scaling as described in Section \ref{sec:method}.  

Predictions obtained via  Eq.~(\ref{eq:GPpredeqns}), with moments
$\mu_{n_F}^b(\mathcal{X})$ and  $\Sigma_{n_F}^b(\mathcal{X})$, evaluated at $\mathcal{X} \equiv \mathbf{X}_{n_M}$ forming $\hat{b}(\cdot)$ via $\mathcal{GP}(D^b_{n_F})$, are shown in the middle and right panels of Figure \ref{fig:MeanAndBias}.
The middle panel shows full $\mu_{n_F}^b(\mathcal{X})$ over depth and DOY, while the right panel shows only the 6m depth slice with 90\% PIs.  Note that $\mu_{n_F}^b(\mathcal{X})$ is much more ``wiggly'' than the surface in the left panel. Focusing on the 6m slice, notice that in early 2021 bias is small in magnitude but may be statistically significant.
Mid-year bias is large and negative with narrow PIs, indicating substantial warm bias.  Finally, later in the year there is  zero/non-significant bias.  

Now consider Figure \ref{fig:PlotForCayelanannotate} (\emph{right}), which shows the bias correction formed as the sum of two GP fits: GLM surrogate $\hat{f}$ and bias correction $\hat{b}$.  One may think of this as (a slice through) the sum of the first two panels in Figure \ref{fig:MeanAndBias}, but that would only account for means.  A more accurate depiction would involve 2d versions of the slices through the full distribution(s) summarized by $\hat{f}$ in Figure \ref{fig:PlotForCayelanannotate} (\emph{right}) and for the bias correction $\hat{b}$ in the right panel of Figure \ref{fig:MeanAndBias}. Mathematically, since $\hat{f}$ and $\hat{b}$ are both MVNs we can follow formulas for sums of MVNs which involve summing the means $(\mu_{n_M}(\mathcal{X}) + \mu^b_{n_F}(\mathcal{X})$) and covariances ($\mathbb{V}\mathrm{ar}\{\mu_{n_M}(\mathcal{X})\} +  \Sigma^b_{n_F}(\mathcal{X})$) and subtracting off cross-covariances.  The bias-corrected surrogate in Figure \ref{fig:PlotForCayelanannotate} omits subtracted cross-covariances due to a negative correlation between the GLM surrogate and bias GP, somewhat overestimating uncertainty. 
However, we do not believe the over-inflation of the total variance negatively affects UQ, as we would prefer conservative UQ over over-confidence of forecasts.

It is important to note that we are not using $\Sigma_{n_M}(\mathcal{X})$, the full predictive variance for $Y(\mathcal{X} \mid D_{n_M})$ from Eq.~(\ref{eq:GPpredeqns}) for the bias corrected (co-) variance.  Instead we are using the nugget-free one (\ref{eq:nonug}) because only the estimated GLM surrogate mean $\boldsymbol{\mu}_{n_M}$  was used to define the observed discrepancies, not actual GLM simulations.  We shall delve into this further in Section \ref{sec:sk}. 
For more details, see \citet[][Sections 5.3.2 \& 8.1.3]{gramacy2020surrogates}.  Observe from Figure \ref{fig:PlotForCayelanannotate} (\emph{right}) that the resulting predictions and PIs are accurate and provide good coverage for the field data.  This is perhaps not surprising as we are predicting in-sample, and the GP (or two GPs in fact) provide a highly flexible non-linear regression.  It is their performance out of sample, for future lake temperatures, which is our focus. 

\section{GP surrogates for forecasting}
\label{sec:method}

There are two challenges to extending the proof-of-concept in Section \ref{sec:back}.  The first is conceptual: pivoting the modeling framework toward forecasts.  We wish to synthesize computer model runs driven by an ensemble of ``plausible weather futures'' so that they may be compared to what actually happened, as observed in the field.  We need a model for how GLM -- driven by NOAA ensembles over extended horizons [Figure \ref{fig:NOAA1}] -- relates to temperatures observed by sensors at FCR.  The second challenge is technological, centering around modeling fidelity and scale. Thirty-day horizons, and 31 NOAA ensemble members over the course of multiple years (and depths) where observational data has been obtained, represent a very large, and heteroskedastic [Figure \ref{fig:NOAA1}], simulation campaign.  Both large-scale GP modeling \citep{heaton2019case}  and accommodating input-dependent variance for stochastic simulation \citep{baker2022analyzing} are on the frontier of the surrogate modeling landscape. Here we describe a setup that provides enough flexibility for our NOAA-GLM at FCR while remaining practical, off-loading work to modern libraries as much as possible, and computationally tractable. 

\subsection{Modeling a forecasting apparatus}
\label{sec:apparatus}

Consider a single horizon of $h$ days into the future, like one day ($h=1$) or one month ($h=30$). We could deploy the setup in Section \ref{sec:back}, except with time ``shifted'' by $t-h$ and with NOAA-simulated environmental variables. More specifically, given NOAA 1--30 day ahead forecasts for each reference date in the training period, GLM will produce 1--30 day ahead lake temperatures at 10 depths. Since the training period is in the past, these faux-future forecasts are what NOAA calls ``hindcasts''.\footnote{Hindcasts are forecasts for past environmental variables pretending the future (also in the past) is unknown.}
We may then compare NOAA-GLM hindcasts to what really happened on that day; i.e.,  we relate the hindcast from time $t-h$ ahead $h$ days to sensor measurements at time $t-h+h \equiv t$.

Figure \ref{fig:NOAA1} shows output from GLM for two reference days $t$, ten depths $d$ and thirty horizons $h$ for 2021. For now, we focus on a single, particular $h$, which amounts to taking a vertical slice through the panels of that figure yielding 31 simulated temperatures at each depth. 
Now average those 31 GLM simulations, separately for each depth, to obtain a single $y^M$-value as output for reference Julian day $t$ and depth $d$.  If this was done for an entire year, we would have the same data setup as in Section \ref{sec:back} -- a single model run for each day and depth in the study period -- but now we would be relating the past (time $t-h$) to the present (time $t$) in a forecasting context.  The data set $D_{n_M}^{(h)}$ would have identical $\mathbf{X}_{n_M}$ but $\mathbf{Y}_{n_M}^{(h)}$ would be different, as indicated by the $(h)$ superscript.  We could then follow Section \ref{sec:back} methods for fitting GP surrogate $\mathcal{GP}(D_{n_M}^{(h)})$ and bias correction $\mathcal{GP}(D_{n_F}^{b(h)})$.  

This can be repeated for each horizon $h$ of interest.  Such a setup has much to recommend it.  It is modular, in that it uses off-the-shelf tools, and it is computationally tractable because each GP is fit to moderately sized training data, keeping the matrix decompositions manageable.  
Such modeling is inefficient statistically because separate, independent fits do not account for correlations across horizon, which are clearly evident in Figure \ref{fig:NOAA1}, both in mean and in variance.  Furthermore, UQ will be less accurate, because through averaging the 31 ensemble GLM simulations, variance information is lost.  Finally, since $t$ represents Julian day, adjustments are required to accommodate multiple years of data.  

Toward a more holistic model, we propose the following.  Let $\mathbf{X}_{n_M}$ represent the matrix of inputs as in Section \ref{sec:back}, but now augmented with a horizon column $h \in \{1,\dots,h_{\max} = 30\}$, an ensemble member column $\xi \in \{1,\dots,31\}$, and a year column $\lambda$.  The $t$ column would be interpreted as the day the forecast was made; the last day for which observational data (i.e., the truth) is known, i.e., $t-h$ from the description above, re-labeled. Columns $\xi$ and $\lambda$ are merely included to index the data, and we shall provide more detail about how these affect modeling momentarily.  We may create such a matrix, i.e., beyond the time and depth columns previously contained in $\mathbf{X}_{n_M}$ from Section \ref{sec:back}, via Cartesian product so that $n_M \leftarrow n_M \times 30 \times 31 \times 3$. Now let $D_{n_M} = (\mathbf{X}_{n_M}, \mathbf{Y}_{n_M})$ collect simulation data relating horizon, ensemble member, time, and depth to GLM-simulated temperature(s).  Supposing we could surrogate model these simulations, we could again follow the recipe in Section \ref{sec:back}, filtering surrogate fits and bias corrections through to forecasts. There are still logistical challenges with $n_M$ in the millions and with heteroskedastic NOAA-GLM runs [Figure \ref{fig:NOAA1}]. 

\subsection{Heteroscedastic surrogate modeling}
\label{sec:sk}
We begin by discarding the ensemble column $\xi$ in $\mathbf{X}_{n_M}$ and treating forecasts from each $\xi$ as independent conditional on time and depth.  This modeling assertion is purely for computational convenience; it is not faithful to the simulation mechanism.  While NOAA ensemble members are independent of one another, each realization evolves in a time-dependent manner.  
Nonetheless, independence aligns with the essence of ensemble modeling \citep{thomas2020near}.
The 31-member NOAA ensemble allows for the measurement of a variety of environmental conditions, whose variability grows over time. We apply these to GLM simulations, as depicted in Figure \ref{fig:NOAA1}. 

This ``modeling hack'' means that each of 31 ensemble forecasts may be treated as \textit{replicates}. \citet{baker2022analyzing} explain that replicates bring two kinds of efficiency.  One is computational: the number of sufficient statistics are fewer -- 31-fold fewer in this instance.  The other is statistical: replicates allow a pure glimpse at variance, without needing to first separate signal from noise.  This is especially important when noise levels are changing along with the mean, in a potentially nonlinear way as in Figure \ref{fig:NOAA1}. The most effective methods for modeling
such nonlinear heteroskedasticity 
\citep[see, e.g.,][]{binois2018practical} leverage replication. \textit{Stochastic kriging} \citep[SK;][]{ankenman2008stochastic}, one of the first of these, is ideal for our setting, in part because it cuts corners relative to  other methods [cited above]; e.g., (a) being moment-based  (rather than likelihood-based); and (b) not jointly modeling the mean and variance process(es).
Indeed, both are valid concerns when the degree of replication is low.  \citet{ankenman2008stochastic} recommend at least ten replicates for each otherwise unique input.  We have 31.
Perhaps more importantly, SK is uniquely situated to interface with our second layer of approximation in Section \ref{sec:vecchia}.

Suppose that simulation outputs $\mathbf{Y}_{n_M}$ correspond to
$n \ll n_M$ unique inputs in $\mathbf{X}_{n_M}$ when ignoring
the $\xi$ column.  In our case $n = n_M / 31$.  Let $\Bar{\mathbf{X}}_n$ denote the matrix of those unique inputs, again without the $\xi$ column, so that $\bar{\mathbf{X}}_n$ is $n \times 4$.  Now, consider each row $\bar{\mathbf{x}}_i$ of $\bar{\mathbf{X}}_n$, and let $y_j(\bar{\mathbf{x}}_i)$ denote the $j^\mathrm{th}$ replicate associated with $\bar{\mathbf{x}}_i$ among the $n_i$ replicates in $\mathbf{X}_{n_M}$.  We have that all $n_i = 31$.  Then, calculate the first two moments for these replicates as
\begin{align}
\Bar{y}_i &= \frac{1}{n_i} \sum_{j=1}^{n_i} y_j(\bar{\mathbf{x}}_i) &
s_i^2 &= \frac{1}{n_i-1} \sum_{j=1}^{n_i} (y_j(\bar{\mathbf{x}}_i) - \Bar{y}_i))^2, \quad\quad \mbox{ for } i=1,\dots,n.
\label{eq:suff}
\end{align}
Finally, store these in $\Bar{\mathbf{Y}}_n$ and $\Bar{\mathbf{S}}_n$,
respectively, to line up with the rows of $\bar{\mathbf{X}}_n$.  
Note that we are dropping the $M$ subscripts with the understanding that we are only using this on the computer model runs. 

SK involves modeling $\Bar{\mathbf{Y}}_n$ and $\Bar{\mathbf{S}}_n$ via GPs rather than operating directly on the original $D_{n_M}$.  It turns out that these $2n$ quantities are not sufficient for $\mathbf{Y}_{n_M}$ but they are close.  \citet{binois2018practical} show that there are $2n + 1$ sufficient statistics.  Nevertheless, \citet{ankenman2008stochastic} argue that inference based on these $2n$ quantities is unbiased asymptotically and minimizes mean-squared error.  Now suppose we wish to predict $\Bar{Y}$ with a GP.  If the degree of replication is uniform, e.g., $n_i = 31$ for all $i$, then we could simply plug the $\bar{\mathbf{Y}}_n$ values, along with $\bar{\mathbf{X}}_n$, into Eq.~(\ref{eq:GPpredeqns}) after solving for MLE hyperparameters.  We can do the same thing for the second moments $\Bar{\mathbf{S}}_n$.  This is the essence of SK with two caveats: (a) uniform replication is rare, although that is the case for NOAA-GLM; (b) it helps to explicitly link the mean and variance processes when making forecasts.  First, fit a GP to $\Bar{\mathbf{S}}_n$ on a transformed scale \citep{johnson2018phenomenological} to respect positivity: $\mathcal{GP}(D^{(v)}_n)$ where $D^{(v)}_n = (\Bar{\mathbf{X}}_n, T(\Bar{\mathbf{S}}_n))$.
For $T(\cdot)$ we prefer the square root to model standard deviations, although a logarithm also works well. 
Then feed $\mathcal{X} \equiv \Bar{\mathbf{X}}_n$ into Eq.~(\ref{eq:GPpredeqns}), and define 
\begin{equation}
\mathbf{S}_n = \mathrm{Diag}(T^{-1} (\mu^{(v)}(\bar{\mathbf{X}}_n))/\mathbf{n}) \quad\quad
\mbox{ where } \quad\quad  \mathbf{n} = (n_1, \dots, n_n).  \label{eq:sepred}
\end{equation}
Several quantities may require explanation.  Predictions $\mu^{(v)}(\bar{\mathbf{X}}_n)$ are transformed back to the original, variance scale.  Dividing these by $n_i$ converts these variances into (squared) standard errors.  Each raw $\sqrt{s_i^2/n_i}$ from Eq.~(\ref{eq:suff}) is a standard error for $\mathbb{E}\{\bar{y}_i\}$, measuring the amount of response information present for each unique input $\bar{\mathbf{x}}_i$. So $T^{-1}(\mu^{(v)}(\bar{\mathbf{x}}_i))/n_i$ is its GP-smoothed analog, borrowing information from nearby standard errors in $\mathbf{x}$-space.  Finally, the diagonal puts these in a matrix representing our otherwise independent (but not identically distributed) structure for noise on the original $\mathbf{Y}_{n_M}$.

Now consider new/forecasting coordinates $\mathcal{X}$ that may be of interest, like for a particular day $t$ and multiple horizons $h$ into the future, or representing hindcasts to compare to sensor measurements discussed further in Section \ref{sec:bias}.  Let $s(\mathcal{X}) = \mathrm{Diag}(T^{-1}(\mu^{(v)}(\mathcal{X})))$ capture these back-transformed variance predictions -- they are the predictive analog of Eq.~(\ref{eq:sepred}).   
Using these, via $\mathbf{S}_n$ and $\mathcal{GP}(D^{(m)})$ where $D^{(m)} = (\Bar{\mathbf{X}}_n, \Bar{\mathbf{Y}}_n)$, the SK equations are
\begin{align}
\mu^{\mathrm{SK}}_n(\mathcal{X}) &= \tau^2 k(\bar{\mathbf{X}}_n, \mathcal{X})^\top (\tau^2 k(\bar{\mathbf{X}}_n) + \mathbf{S}_n)^{-1} \boldsymbol{\Bar{Y}}_{n} \label{eq:SKmu} \\
\Sigma^{\mathrm{SK}}_n(\mathcal{X}) &= \tau^2 (k(\mathcal{X}) + s(\mathcal{X})) - \tau^2 k(\bar{\mathbf{X}}_n, \mathcal{X})^\top (\tau^2 k(\bar{\mathbf{X}}_n) + \mathbf{S}_n)^{-1} \tau^2k(\bar{\mathbf{X}}_n, \mathcal{X}).  \label{eq:SKsigma}
\end{align}
Observe in Eq.~(\ref{eq:SKmu}) that augmenting the diagonal of $k(\bar{\mathbf{X}}_n)$ with $\mathbf{S}_n$ results in a scaled nugget-like term, by analogy to $\tau^2 g$ in Eq.~(\ref{eq:GPpredeqns}).  It facilitates a signal-to-noise trade-off that is unique to each training data element. 
Smoothed (squared) standard errors $\mathbf{S}_n$ naturally downweight large variance/low replication training data.  
It may be shown that $\mu^{\mathrm{SK}}_n(\mathcal{X}) = \mu_{n_M}(\mathcal{X})$ from Eq.~(\ref{eq:GPpredeqns}), identically, 
which could represent a huge computational savings when $n \ll n_M$.

\begin{figure}[ht!]
\centering 
\vspace{-0.25cm}
\includegraphics[scale=0.18,trim={30 10 40 0}]{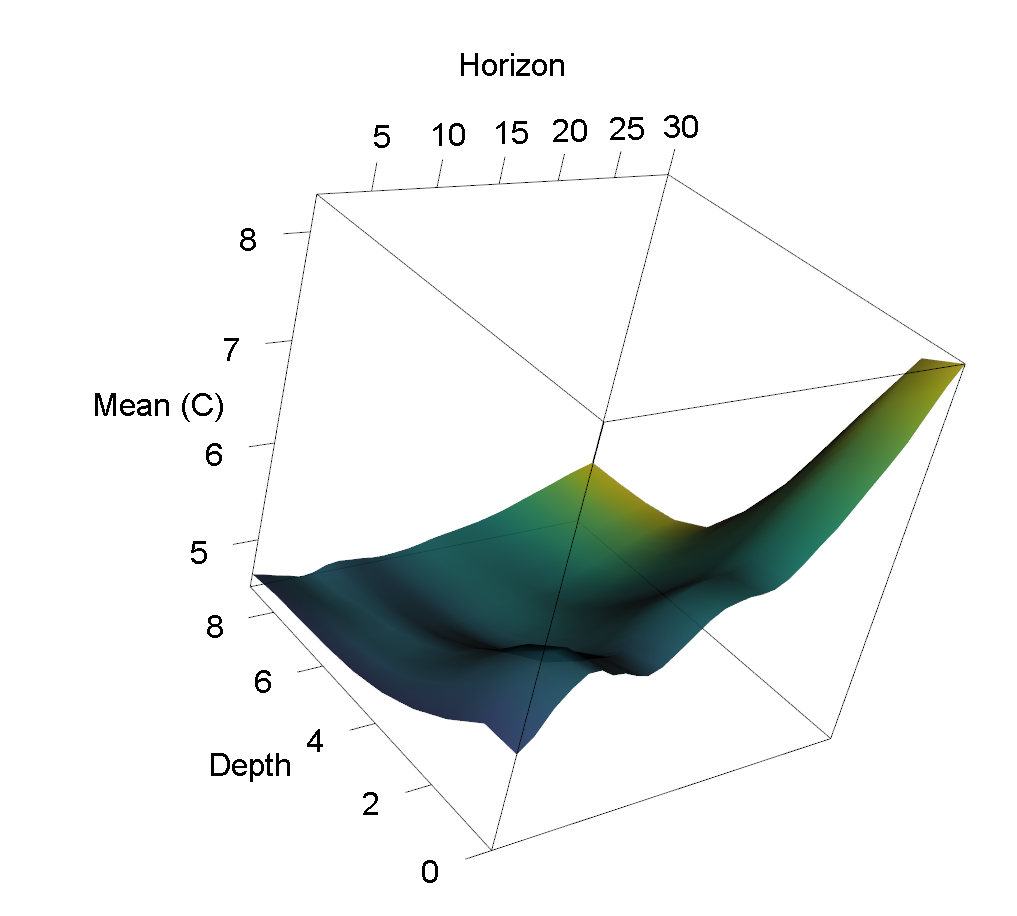} 
\hspace{0.1cm}
\includegraphics[scale=0.18, trim={40 10 20 0}]{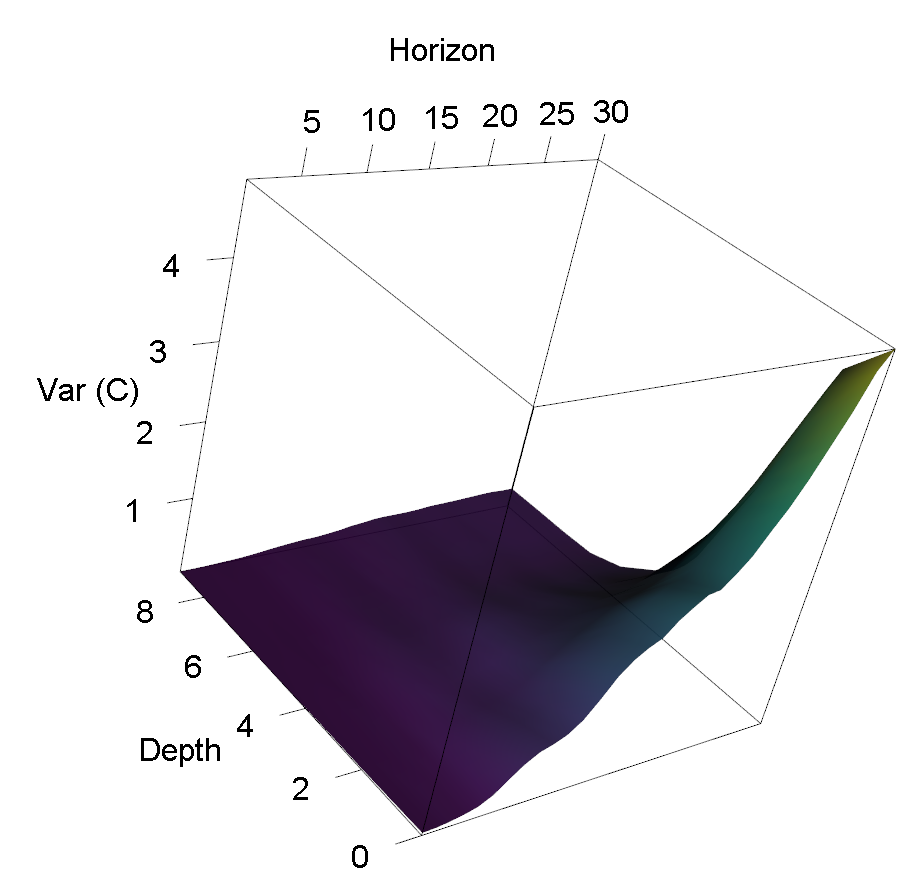} 
\hspace{0.1cm}
\includegraphics[scale=0.21, trim={0 0 0 10}]{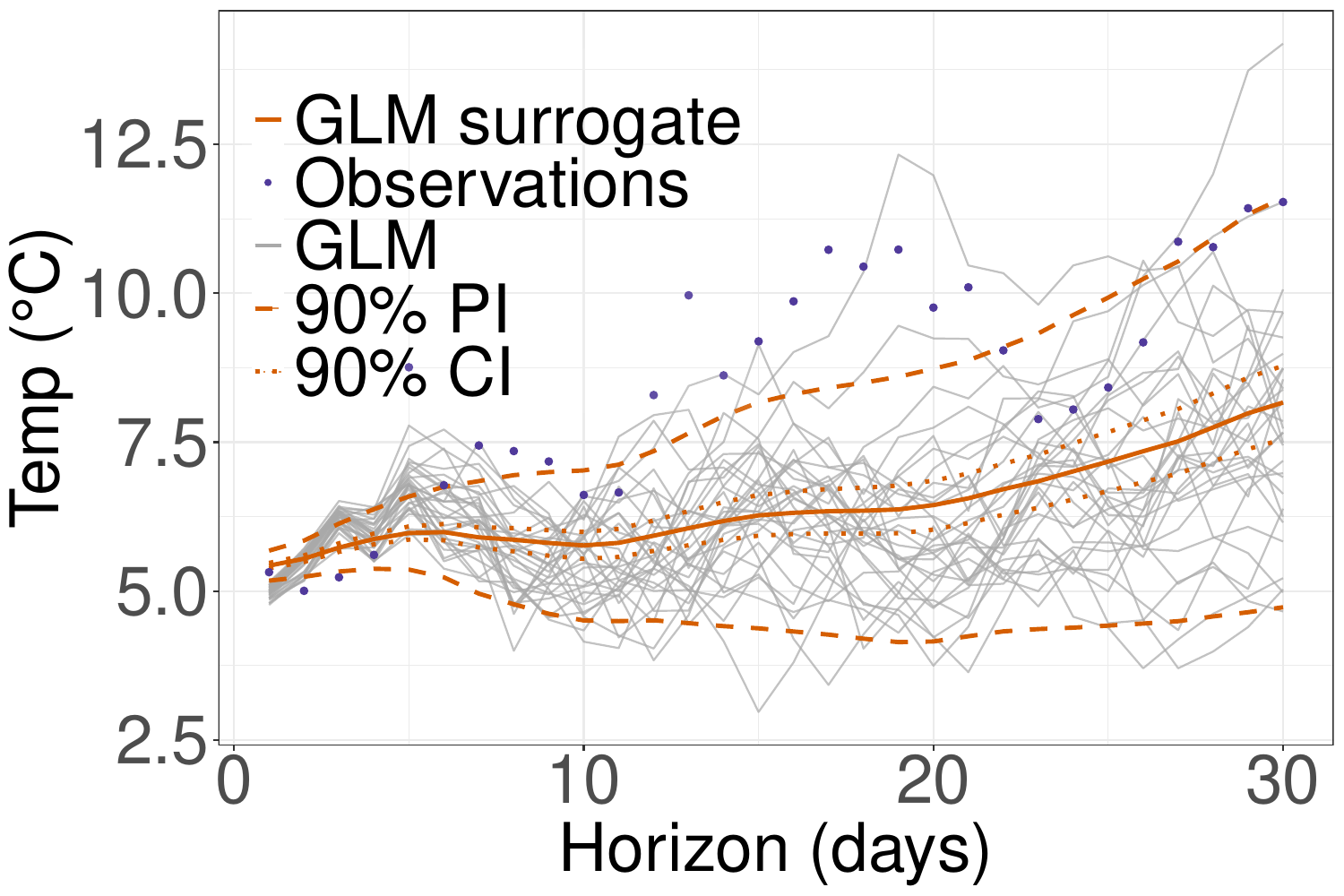}
\caption{2d slices of predictive mean  $\mu^{\mathrm{SK}}_n(\mathcal{X})$ (\textit{left}) and variance $\sigma^{\mathrm{SK}}_n(\mathcal{X})^2$ (\textit{middle}) from the (heteroscedastic) GLM surrogate over depth (m) and horizon (days). Together, these summarize a hindcast for 2022-02-19.  The {\em right}
panel shows a 1m-depth slice with GLM ensemble forecasts (light gray lines), SK surrogate (darker, solid line) 90\% CIs and PI (dotted and dashed lines). Observations are overlaid as darker dots.}\label{fig:heteroex2} 
\end{figure}

To illustrate, consider Figure \ref{fig:heteroex2} which is derived from fits of two GPs, for mean and variance, to a simulation campaign spanning three years over a thirty-day horizon. The data represented in the plot comprises a 2d slice of predictions from those two fits corresponding to a forecast beginning on 2022-02-19, offering a surrogate for the NOAA-GLM hindcast beginning on that date.  Focusing first on the left panel, showing  $\mu^{\mathrm{SK}}_n(\mathcal{X})$ from Eq.~(\ref{eq:SKmu}), observe that temperature is decreasing with increasing depth, and generally increasing over horizon, most prominently at the surface (low depths). Now looking at the middle panel providing $\sigma^{\mathrm{SK}}_n(\mathcal{X})^2 = \mathrm{Diag}(\Sigma^{\mathrm{SK}}_n(\mathcal{X}))$ we see a similar relationship in the spread of possible temperatures.  Variance is much lower at lower depths, even at longer horizons, because these depths are more insulated from weather at the surface.  The right panel of the figure shows a 1d slice at 1m depth.  Gray lines indicate raw NOAA-GLM simulations (i.e., a subset of $D_{n_M}$).  Solid dots indicate sensor observations. 
Observe that the 90\% PIs (dashed lines), derived from $\mu^{\mathrm{SK}}_n(\mathcal{X})$ and $\sigma^{\mathrm{SK}}_n(\mathcal{X})^2$,  have appropriate coverage.  Dotted lines indicate  90\% CIs on the mean that are derived by eliminating the $s(\mathcal{X})$ term in Eq.~(\ref{eq:SKsigma}), as an analogue to Eq.~(\ref{eq:nonug}).   It is this squared standard error which is crucial to our bias-correcting analysis described in Section \ref{sec:bias}. Finally, observe that NOAA-GLM has a cold bias compared to the field data. 

\subsection{GP surrogates and scalability}
\label{sec:vecchia}
The surrogates behind Figure \ref{fig:heteroex2}  were actually constructed via GP approximation, which has been an active field of late.  
The so-called scaled Vecchia approximation \citep[SVecchia;][]{katzfuss2022scaled} seems
particularly well-suited to our setting because it allows coordinate-wise lengthscale estimation and scales in flops that are quasi-linear in $n$.  Crucially, an open-source implementation is provided for {\sf R}: \url{https://github.com/katzfuss-group/scaledVecchia} via \texttt{GpGp} \citep{guinness2018permutation} and \texttt{GPVecchia} \citep{Katzfuss2020GPVecchia} packages.  We provide a review of the Vecchia approximation, and some implementation details in Section A.1 of the Supplementary Material \citep{HolthuijzenSupplement}.

Vecchia has never been used, and nor to our knowledge has any other large-scale GP approximation, to model input-dependent variances, like through SK (Section \ref{sec:sk}). 
Fitting second moments $\mathcal{GP}(D_n^{(v)}$) in order to get a predictor for $T( \mathbb{V}\mathrm{ar}\{Y(\mathbf{\mathcal{X}})\} )$ is straightforward, because this amounts to feeding in (transformed) $s_i$-values (Eq. \ref{eq:suff}) as responses along with unique inputs $\bar{\mathbf{x}}_i$.  The visual in the right panel of Figure \ref{fig:heteroex2} shows
$\mu^{(v)}_n(\mathcal{X})$.  Using Vecchia to fit the mean GP conditional on those variances (left panel of Figure \ref{fig:heteroex2}) is harder because the {\sf R} library we use does not accommodate a vector of nuggets, which comprises $\mathbf{S}_n$ in Eqs.~(\ref{eq:SKmu}--\ref{eq:SKsigma}).  
We implemented the following workaround instead.

We begin by fitting first moments with an otherwise ordinary Vecchia
approximated GP, i.e., $\mathcal{GP}(D_n^{(m)})$,  where $D_n^{(m)} = (\Bar{\mathbf{X}}_n, \Bar{\mathbf{Y}}_n)$. 
Let $\mu^{(m)}_n(\mathcal{X})$ denote the predictive mean of $\mathcal{GP}(D_n^{(m)})$. Note that $\mu^{(m)}_n \ne \mu^{\mathrm{SK}}_n$ from Eq.~(\ref{eq:SKmu}) because the former utilizes a covariance structure with a single nugget, i.e., $\tau^2(k(\bar{\mathbf{X}}_n) +  g\mathbb{I}_n)$ following Eq.~(\ref{eq:GPpredeqns}), and the latter uses
$\tau^2k(\bar{\mathbf{X}}_n) + \mathbf{S}_n$.  
However, they are equal in expectation because GPs furnish unbiased predictors \citep[see, e.g.,][]{santner2018design}.
Nevertheless, the two can be quite
different if the degree of replication is high because of the $\mathbf{n}$-vector in the denominator of Eq.~(\ref{eq:sepred}).  Specifically, if the replication degree varies significantly from one location to another, for instance $n_i \gg n_j$ for $\bar{\mathbf{x}}_i$ and $\bar{\mathbf{x}}_j$, then the predictions from $\mu^{\mathrm{SK}}_n(\mathbf{x})$ for $\mathbf{x}$ in the vicinity of $\bar{\mathbf{x}}_j$, the site with fewer replicates, could be considerably more precise than $\mu^{(m)}_n(\mathbf{x})$, which does not know to place less trust in $\bar{y}_j$ than in other locations, such as near $\bar{\mathbf{x}}_i$. However, in our setting, we have uniform replication, with all $n_i = 31$.  In a study described in Section A.3 and summarized in Figure A1 of the Supplementary Material \citep{HolthuijzenSupplement}, we find there is essentially no difference between $\mu^{(m)}_n$, via Vecchia, and $\mu^{\mathrm{SK}}_n$. 

Approximating what SK provides for the variance of the first moment, $\Sigma^{\mathrm{SK}}_n(\mathcal{X})$ in Eq.~(\ref{eq:SKsigma}), with Vecchia, follows a similar thread.  Whereas our development of the mean leverages the unbiased nature of Gaussian expectations, second moments follow standard calculations of Gaussian variances.  Crucially, we do not need the entire covariance structure $\Sigma^{\mathrm{SK}}_n(\mathcal{X})$ for bias
correction in Section \ref{sec:bias}; we need only the diagonal.  Those ``variances of the mean'' are Gaussian standard errors, 
i.e., $\mathbb{V}\mathrm{ar}\{Y(\mathbf{\mathbf{x}})\}/n$, again 
leveraging uniform replication: $n_i = 31$.  Our Vecchia fit to second moments provides an \emph{estimate} of $\mathbb{V}\mathrm{ar}\{Y(\mathbf{\mathbf{x}})\}$ so all we need to do is divide this quantity by $n_i$, for any $i$, to get an estimate of the variance of the mean. Importantly, we prefer a more conservative application of this in an SK-approximating context to avoid under-covering out-of-sample. We found that using the 95th quantile of the predictive mean obtained from $\mathcal{GP}(D_n^{(v)})$, notated as $\mu^{(v)95}_n(\mathbf{x})$, added to the variance of the first moments' fit ($\sigma^{2m}_n(\mathbf{x})$) provided sufficient coverage. That is, we use
\begin{align}
\label{eq:vsk}
\hat{\mu}_n^{\mathrm{SK}}(\mathbf{x}) &= \mu_n^{(m)}(\mathbf{x}) & \mbox{ and } &&
\hat{\sigma}^{\mathrm{SK}}_n(\mathbf{x})^2 &= \mu^{(v)95}_n(\mathbf{x})/n_i + 
\sigma^{2(m)}_n(\mathbf{x}).
\end{align}
Beyond missing off-diagonal covariance terms, this is
an inefficient estimator compared to the ideal SK quantity in
Eq.~(\ref{eq:SKsigma}) because it does not borrow strength nearby
in the input space.  (This is what the term $- \tau^2 k(\mathbf{X}_n, \mathcal{X})^\top (\tau^2 k(\mathbf{X}_n) + \mathbf{S}_n)^{-1} \tau^2k(\mathbf{X}_n, \mathcal{X})$ in that
expression is doing.)  So in $\hat{\sigma}^{\mathrm{SK}}_n(\mathbf{x})^2$ 
we are getting over-inflated, conservative UQ because the Vecchia software
we are using does not provide a full covariance. Note that $\hat{\sigma}^{\mathrm{SK}}_n(\mathbf{x})^2$ from Eq. (\ref{eq:vsk}) will give rise to CIs; however, if division by $n_i$ is omitted, one may obtain PIs. The construction of PIs in this manner is only valid for the GLM surrogate without bias-correction, as will be explained in the next section. 

Sections A.2--A.3 of the Supplementary Material \citep{HolthuijzenSupplement} provide two examples showing that this approach achieves the desired results in a benchmark/toy example and on GLM runs, respectively.  These are intended to augment other visuals in the paper, which are {\em all} derived from Vecchia approximations.

\subsection{Modeling forecast bias}
\label{sec:bias}
Now, let $\hat{\mu}_n^{\mathrm{SK}}(\mathbf{x})$ and $\hat{\sigma}^{\mathrm{SK}}(\mathbf{x})^2_n$ denote the moments of that NOAA-GLM surrogate (Eq. \ref{eq:vsk}), combining Sections \ref{sec:sk}--\ref{sec:vecchia},
but fit to the entire campaign of runs across all
31 ensemble members and all training days/horizons/depths under study: ($\mathbf{X}_{n_M}, \mathbf{Y}_{n_M}$), and $n_M \approx$ 8 million-odd runs.  Providing $\mathbf{x} \in \mathcal{X}$
representing any time point(s) of interest (future, present or past), including depths and horizons of interest, yields a prediction (or forecast) capturing uncertainty in GLM dynamics driven by NOAA ensemble hindcasts.  
Recall from Section \ref{sec:GPlaketemps} that estimating a bias correction requires measuring discrepancy
between computer model predictions and field data observations,
forming a data set $D^b$ that can be used to train a GP.  Our
field data are a single set of observations $\mathbf{Y}^F_{n_F}$ 
in time and depth, coded in $\mathbf{X}_{n_F}$.  In Section \ref{sec:GPlaketemps} we had $\mathbf{X}_{n_F} = \mathbf{X}_{n_M}$, so $\mathbf{Y}^F_{n_F}$ could easily be lined up with predictions
$\boldsymbol{\mu}_{n_M}$ to measure discrepancies (\ref{eq:Calcbias}).
  However, now our set of GLM inputs (and outputs) is expanded
[Section \ref{sec:apparatus}] to encompass 
horizon, which is not relevant to our field sensor measurements. Field data and surrogate are out of alignment.

Let $\bar{\mathbf{Y}}^F_{n_F}$ denote field data responses ``matching'' unique computer model
inputs.  Building this vector requires 
inserting multiple copies $\mathbf{Y}^F_{n_F}$ into $\bar{\mathbf{Y}}^F_n$ in such a way that they line up with the simulation parameters corresponding to the time/depth/horizon parameters in $\bar{\mathbf{X}}_n$.  For example, suppose $\mathbf{x}_i^\top$, residing in the $i^{\mathrm{th}}$ row of $\bar{\mathbf{X}}^F_{n_F}$ records a particular time $t_i$ and depth $d_i$ measurement of lake temperature.  Then, place $y_i$, the $i^\mathrm{th}$ entry of the vector $\mathbf{Y}^F_{n_F}$, in all 30 slots $\bar{\mathbf{Y}}_{n_F}^F$ so that they match up with the $\mathbf{x}$-tuple $(t_i - h_i, d_i, h_i)$ in $\bar{\mathbf{X}}_n$.  Then, let $D^b_{n} = (\bar{\mathbf{X}}_n, \bar{\mathbf{Y}}^F_{n_F}
 - \hat{\mu}_n^{\mathrm{SK}}(\mathbf{X}_n))$ denote discrepancies (\ref{eq:Calcbias}) and fit $\mathcal{GP}(D^b_n)$ to characterize a bias correction over time/depth/horizon.
The size of $n$ again makes full GP fitting prohibitive, so
we use a Vecchia approximation.  However, we do not have the
luxury of replication, so we must opt for an ordinary, homoskedastic GP fit. Let $\mu_{n}^b(\cdot)$ and $\sigma_n^{2b}(\cdot)$ denote the moments of the GP fit to $\mathcal{GP}(D^b_n)$, overloading our earlier Section \ref{sec:GPlaketemps} notation.  A GP surrogate with bias-correction (GPBC) may be formed by chaining the two sets of predictive moments together:
\begin{align}
\mu_n(\mathcal{X}) &\equiv \hat{\mu}_n^{\mathrm{SK}}(\mathcal{X})+ \mu_n^b(\mathcal{X}) \label{eq:bc} & \mbox{ and } &&
\sigma_n^2(\mathcal{X}) &\equiv \hat{\sigma}_n^{\mathrm{SK}}(\mathcal{X})^2+ \sigma_n^{2b}(\mathcal{X}).
\end{align}
Recall that $\hat{\mu}_n^{\mathrm{SK}}(\mathcal{X})$ is the predictive mean of the first moments' fit $\mathcal{GP}(D_n^{(m)})$.
As in Section \ref{sec:GPlaketemps}, $\sigma_n^2(\mathcal{X})$ over-estimates the variance of the sum of two GP variances because it is missing a negative covariance term. In contrast to the discussion
in Section \ref{sec:GPlaketemps}, we cannot utilize full predictive
covariance matrices $\Sigma(\mathcal{X})$ because the Vecchia implementation we use only provides the diagonal, which is what we are denoting as $\sigma_n^2(\mathcal{X})$.  The most useful settings for $\mathcal{X}$ indicate hindcasts ($\mathcal{X}$ depicting a historical set of forecasting days, horizons, and depths) and forecasts ($\mathcal{X}$ denoting the current day, a horizon of the next thirty days, and all depths of interest). 

To summarize, the NOAA-GLM surrogate arises from fitting GPs to first and second moments,  $\bar{\mathbf{Y}}_n$ and $\bar{\mathbf{S}}_n$, respectively, via methods described in Sections \ref{sec:sk} and \ref{sec:vecchia}. The predictive mean, $\hat{\mu}_n^{\mathrm{SK}}(\mathcal{X})$ is subtracted from observations  $\bar{\mathbf{Y}}^F_n$ to form a dataset of discrepancies $D^b_n$, to which a homoscedastic GP is fitted. Finally, the GP surrogate with bias-correction (GPBC) is obtained by adding predictive moments from both the surrogate and bias correction.

\begin{figure}[ht!]
\includegraphics[scale=.6, trim={0 5 0 0}, clip]{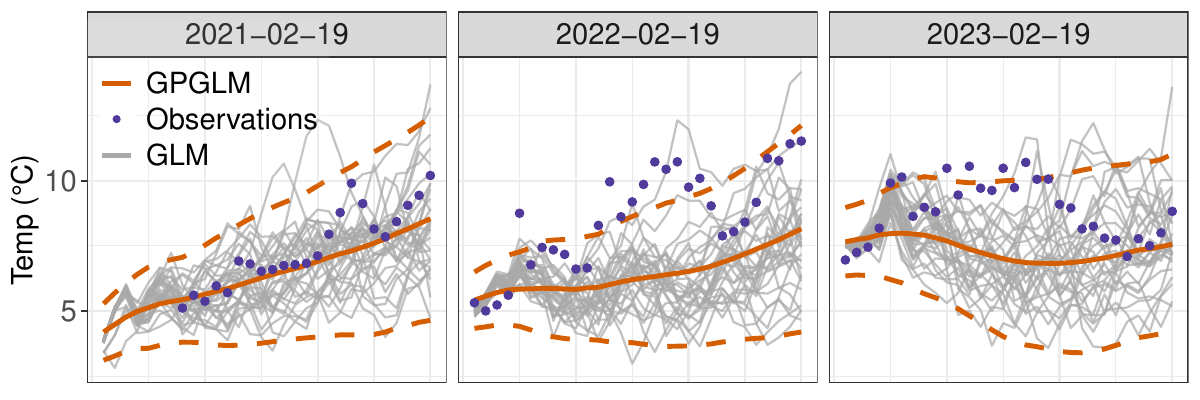}
\includegraphics[scale=.6, trim={0 8 5 6}, clip]{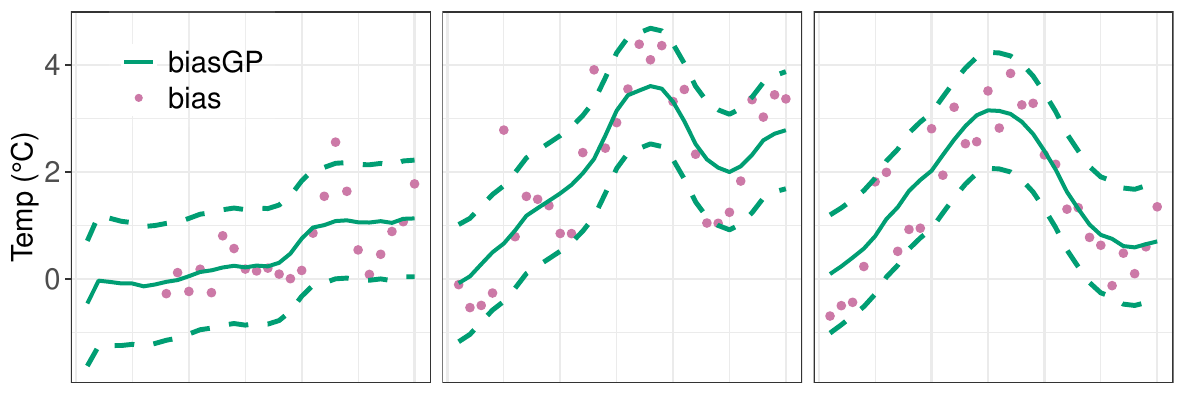}
\includegraphics[scale=.6, trim={0 0 0 5}, clip]{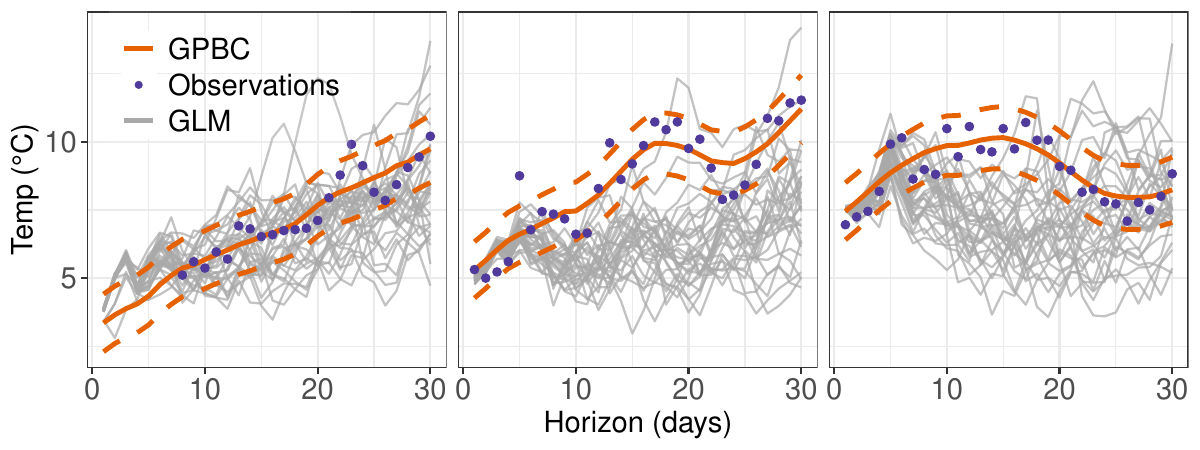}
\caption{NOAA GLM forecasts (thin gray lines) starting on Feb. 19 for years 2021, 2022, and 2023 at a depth of 1m. Sensor measurements (dark points) are overlaid. Top row: In-sample predictions and 90\% PIs from the GLM surrogate (GPGLM, thick  lines). Middle row: Means and 90\% PIs for the bias GP (biasGP) with bias values overlaid as points. Bottom row: In-sample forecasts (predictive means) and 90\% PIs for the \emph{bias-corrected} GLM surrogate (GPBC, thick lines) with observations overlaid as dark points.}
\label{fig:BCnoBCtrainperiodNOOGP}
\end{figure}

To illustrate, Figure \ref{fig:BCnoBCtrainperiodNOOGP} shows examples of 1m-depth ``slices'' of hindcasts via quantities defined above starting on Feb. 19, 2021 (left), 2022 (middle) and 2023 (right). Each thin gray line in the top row is a NOAA-GLM hindcast over a 30-day horizon.  Thicker lines summarize the computer model surrogate and 90\% PIx (via $\hat{\mu}_n^{\mathrm{SK}}$ and $\hat{\sigma}_n^{\mathrm{SK}}$ without dividing by $n_i$), denoted as GPGLM.
Darker dots show the $\bar{\mathbf{Y}}^F_{n_F}$ corresponding to that date/depth and each horizon. Discrepancy data $D^b$, derived from the $\hat{\mu}_n^{\mathrm{SK}}$ and  $\bar{\mathbf{Y}}^F_{n_F}$, are shown in the middle panel.  Observe that in all three panels/years, these fits exhibit a cold bias because the bias correction is positive. 
In 2021 correction from zero might not be statistically significant.  In 2022 and 2023 it is substantial, highly non-linear, and error-bars do not cover zero over most of the 30-day horizon.  
 
The bottom row of the figure shows how Eq.~(\ref{eq:bc}) is used to combine the quantities in the top and middle row, showing the bias fit, to obtain a bias-corrected surrogate (GPBC), for which the mean and variance are given by Eq. (\ref{eq:bc}).  Observe that in all three years this is a good approximation to sensor measurements.  Note that while PIs for the surrogate in the top panels exhibit a trumpet-like shape over increasing horizon, those in the bottom panels no longer have this pronounced shape. Note that PIs for GPGLM use $\hat{\sigma}^{\mathrm{SK}}_n(\mathbf{x})^2$ without division by $n_i$ while GPBC leverages $\hat{\sigma}^{\mathrm{SK}}_n(\mathbf{x})^2$ with division by $n_i$ as written in Eq. (\ref{eq:vsk}). The total variance of GPBC is obtained by adding $\hat{\sigma}^{\mathrm{SK}}_n(\mathbf{x})^2$ to the predictive variance from the bias GP $(\sigma_n^{2b}(\mathcal{X}))$, shown in Eq.~(\ref{eq:bc}). Because $\hat{\sigma}^{\mathrm{SK}}_n(\mathbf{x})^2 \ll \sigma_n^{2b}(\mathcal{X})$ from the (homoskedastic) bias GP, the overall predictive variance of GPBC appears homoscedastic.

\begin{figure}[ht!]
\centering
\includegraphics[scale=0.6]{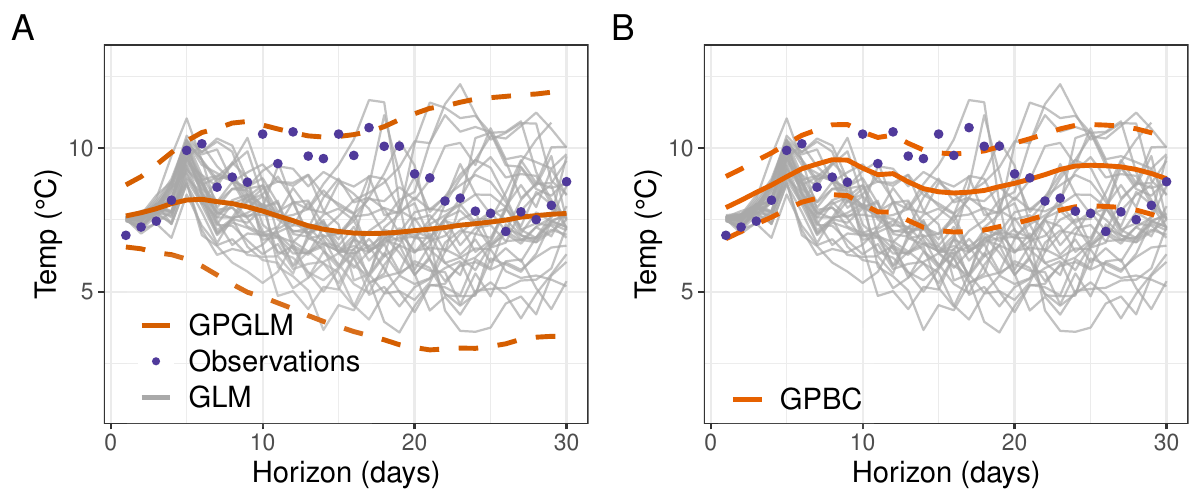}
\caption{NOAA GLM forecasts (gray lines) starting on Feb. 19 for year 2023 at a depth of 1m; sensor measurements are shown as darker dots. Panel A shows the in-sample prediction from the GLM surrogate as depicted in \ref{fig:BCnoBCtrainperiodNOOGP} (\emph{right panel}). Panel B shows the out-of-sample forecast from the \emph{bias-corrected} GLM surrogate.}
\label{fig:BCnoBCTestperiod}
\end{figure}  

Figure \ref{fig:BCnoBCTestperiod} shows a similar pair of views, but now in a forecasting context corresponding to the right-most column of Figure \ref{fig:BCnoBCtrainperiodNOOGP} (2023).  The gray lines from the GLM were used to build the surrogate (along with simulations for all other days/depths in the past), but the field data (dark dots) are not observed.  Thus, bias correction is based only on patterns learned on \emph{historical} data. 
Although the overall predictive quality in Figure \ref{fig:BCnoBCTestperiod}  is
lower than the bottom row of Figure \ref{fig:BCnoBCtrainperiodNOOGP}, because the former is in-sample and the latter is out-of-sample, it is clear the GPBC performs well.  Visually, the bias correction is effective (panel B) compared to the surrogate on its own (panel A).  This can also be quantified as follows:
GPBC achieves an RMSE of 1.24$^\circ$C compared to 1.86$^\circ$C GPGLM on its own. Note that while PI coverage is lower for GPBC for this particular example, the width of the PIs of GPBC (2.23) is less than half of the PI width of the surrogate on its own (5.14).

\section{Real-time forecasting}
\label{sec:realtime}

We now have all of the ingredients in place to build a real-time forecasting system combining simulated and observed temperatures.  
Here we elaborate on the configuration of our system, with an emphasis on the principal variation based on Vecchia/SK, as described in Section \ref{sec:method}, while also accommodating more straightforward alternatives earlier.  We then describe a variation that allows forecasts to react more dynamically to recent observational and simulation data, and to data from the most similar previous years.


\subsection{Iterative forecasting}
\label{sec:iterative}

Our forecasting framework consists of: an initial training stage, in which many historical NOAA-GLM simulations are used to fit a GP surrogate  $\mathcal{GP}(D_{n_{M}})$ and bias correction $\mathcal{GP}(D^b_n)$; and an iterative stage, in which new data are downloaded and a new forecast is generated from updated surrogates.  Here we are using $\mathcal{GP}$ generically, since many variations are possible. The initial training phase begins with obtaining NOAA-GLM blue hindcasts for all reference dates in the historical training period. A surrogate fit to these data provides a statistical abstraction of NOAA-GLM simulations over the historical period of time. Observed residuals from sensor data are used to estimate a bias correction.
The iterative phase begins with downloading NOAA ensemble forecasts over the next thirty days. Next, these are fed as driver data to the GLM, augmenting the simulation campaign from the training phase, and updating the surrogate and bias correction accordingly. The bias-corrected surrogate can then be used to make lake temperature forecasts over that thirty-day period.  The process repeats the next day after observing actual temperatures on the current day from sensors at each depth. Additional detail, along with a diagrammatic depiction [Figure B3] is provided in Section B of the Supplementary Material \citep{HolthuijzenSupplement}.

\subsection{Incorporating annual variability into forecasts}
\label{sec:phi}

The inputs to our surrogate modeling and bias correction framework are Julian day $t \in \{1,\dots, 365\}$, depth $d \in \{0,\dots,9\}$, and horizon $h \in \{1,\dots,30\}$.  A consequence of this choice, especially with regard to the 
time input $t$, is that our forecasts will be identical for any day $t$ in any year.  We have a ``day of year'' model.
As we gather more data, forecasts will have more conservative UQ because they will have been trained on a wider diversity of observations across year for that day, and days nearby in time.  In other words, they will better capture climatological variability, but they will also be less reactive to current weather patterns and other environmental variables. 

\begin{figure}[ht!]
\centering
\includegraphics[scale=0.45]{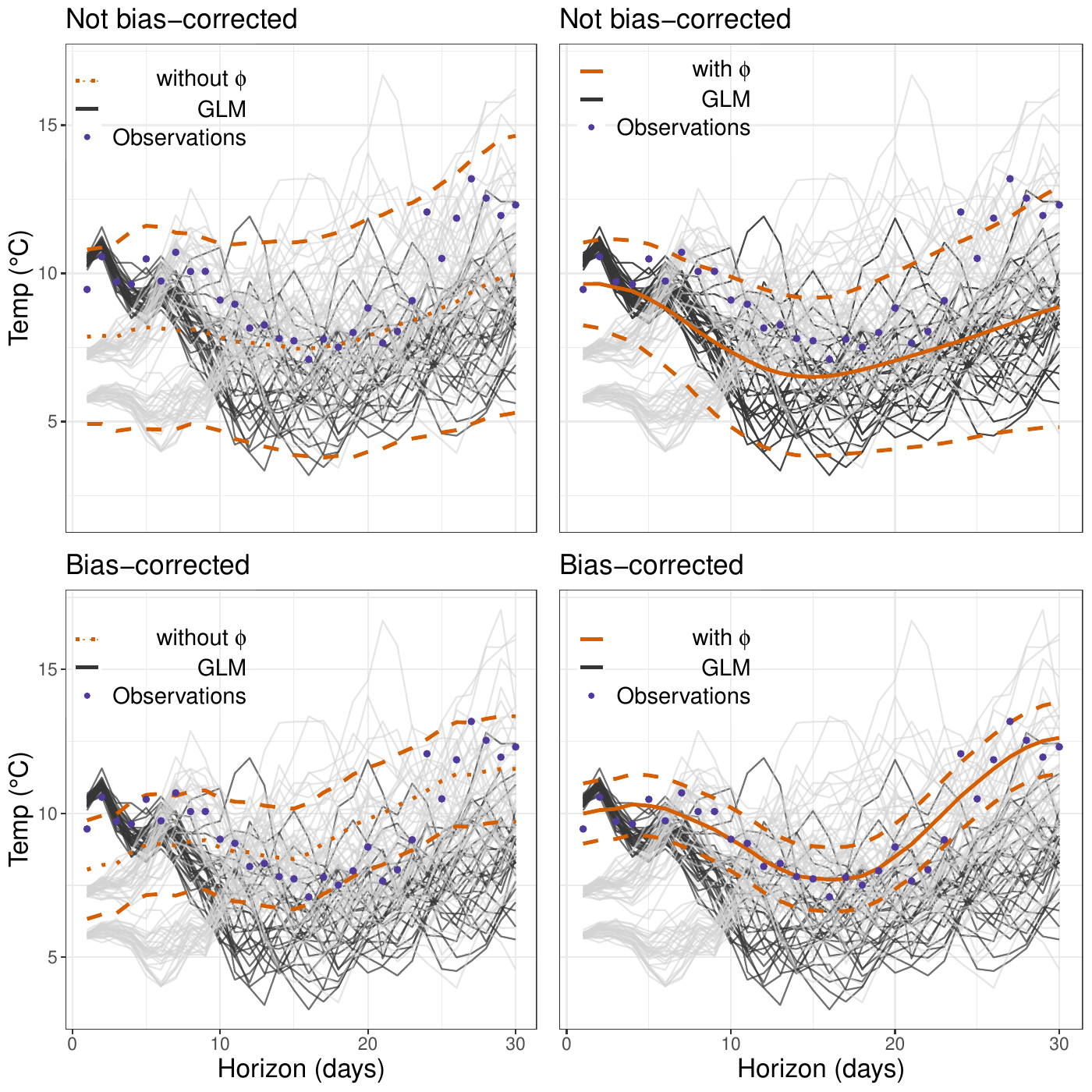}
\caption{In-sample forecasts for March 1, 2023 models without $\phi$ (``without $\phi$''; left panels, dotted lines) and those with $\phi$ (``with'' $\phi$); right panels, solid lines) at 1m depth; 90\% PIS are denoted by dashed lines, Sensor measurements as dark dots. GLM forecasts for March 1, 2023 are in dark gray; previous years in light gray.}
\label{fig:GPBCphivsGPBC}
\end{figure}

Toward obtaining more dynamic forecasts, we introduce a fourth input variable, $\phi$, whose behavior is illustrated in Figure \ref{fig:GPBCphivsGPBC}. To illustrate how the ``with $\phi$'' model differs from its ``without $\phi$'' counterpart, consider the left column of Figure \ref{fig:GPBCphivsGPBC}, which shows examples of predictions over a 30-day horizon for March 1, 2023 at depth 1m from the surrogate only (top panel) and after bias correction (bottom). 
In contrast to previous visuals, we now show all three sets of NOAA-GLM simulations available up until this time point: today's (March 1 2023) in black and the previous two years (March 1 in 2022 and 2021) in gray.  Observe that the forecasts do a poor job of tracking today's black simulations.  In the top row, forecasts from the surrogate track the corpus of simulations, whereas in the bottom row their spread is narrowed by the bias correction, but in an overly conservative fashion.  

The right column of Figure \ref{fig:GPBCphivsGPBC} explores the effect of adding $\phi$ as a predictor. Observe how, in contrast with the left column without $\phi$, the GLM surrogate without a bias-correction (top) provides a good fit to NOAA-GLM forecasts (black lines). When bias-correction is added (bottom), prediction intervals narrow considerably compared to the sensor measurements (dark dots).

With $\phi$ we aim to capture the unique ``state'' for day $t$ and year $\lambda$  [first introduced in Section \ref{sec:apparatus}], via the following calculation
\begin{equation}
    \phi(t,\lambda) = \frac{1}{5} \sum_{i = 0}^{4} \left ( \frac{\mathbf{Y}^F_{n_F}(t-i, d=0, \lambda) + \mathbf{Y}^F_{n_F}(t-i, d=1, \lambda) }{2}\right ),
    \label{eq:phi}
\end{equation}
i.e., average sensor-measured temperatures near the surface (depths $d$ = 0 and 1 meters) over the previous five days. Temperatures near the surface are more variable than those at deeper layers, and are therefore a good indication of weather conditions. We augment design matrices $\bar{\mathbf{X}}_n$ 
and $\bar{\mathbf{X}}_{n_F}^F$ to include $\phi(t,\lambda)$ matching up with $t$ and $\lambda$. 
Surrogate modeling and bias correction, as described in Sections \ref{sec:sk}--\ref{sec:bias}, are not affected by this change excepting  higher input dimension and additional lengthscale(s).  In particular, responses $\bar{\mathbf{Y}}_n^M$ and  $\mathbf{Y}^F_{n_F}$ remain as before. 

We generally find better results with the addition of $\phi$,
as demonstrated in Section \ref{sec:results}.  
Consequently, we shall going forward take the $\phi$ version to be the main/default modeling variation unless we mention that modeling ``without $\phi$'' was entertained as a contrast.  In fact, all previous visuals in the paper used the version with $\phi$, because they looked better.  
Eq.~(\ref{eq:phi}) represents only one of many ways that a year-differentiating state could be calculated.  It could even be treated as a latent variable [see Section E of the Supplementary Material \citep{HolthuijzenSupplement}].

\section{Out of sample results}
\label{sec:results}

Here we lay out evaluation metrics and comparators and then describe an out-of-sample exercise designed to explore the relative merits of GPBC forecasts.

\subsection{Validation and benchmarking}
\label{sec:validate}

When new sensor measurements come in at midnight we have the opportunity to assess the accuracy of yesterday's forecasts, as predicted on all thirty previous days at varying future horizons.  We evaluate (1) RMSE; (2) a proper log score  \citep[Eq.~(25) of][]{gneiting2007strictly}; (3) confidence interval width and (4) coverage.  For details see Section C of the Supplementary Material \citep{HolthuijzenSupplement}. Smaller RMSEs and larger scores are preferred. 

As comparators we consider the following four methods. Each is simplification of our flagship GPBC (with $\phi$) approach, missing one or more of its building blocks. 
\begin{itemize}
\item[\underline{GPGLM:}] a GP surrogate for GLM without bias-correction (GPGLM), but otherwise using all predictors (including $\phi$) and Vecchia/SK for fast heteroskedastic modeling. 
\item[\underline{OGP:}] a heteroscedastic GP fitted to observed sensor measurements only, using Julian day of year and depth as inputs (no $\phi$).  
These forecasts simply reflect a ``typical'' year based all previous years' measurements.  Importantly, GLM's short-term forecasts are not used at all.  In this way, OGP represents a ``climatological'' model, as its predictions reflect typical seasonality and general trends. Since the sensor data set is smaller ($365\times 10$/year) we do not need Vecchia, so we opted for a heteroskedastic GP fit from {\tt hetGP} \citep{binois2021hetgp} via {\tt mleHetGP} for joint inference for first and second-moment GPs.
\item[\underline{GLM:}] raw 1--30 day-ahead NOAA-GLM forecasts in the testing time period.   We calculated empirical means and standard deviations over all 31 ensemble members for all combinations of reference date, depth, and horizon in the test time period. 
\item[\underline{GPBC} without $\phi$:] a bias-corrected surrogate for GLM where the input $\phi$ omitted. Like OGP, forecasts are based only on historical information. However, it does incorporate NOAA-GLM hindcasts for the current day, so it is not purely climatological.
\end{itemize}

Each benchmarking comparator offers a glimpse into the potential benefit, or hazard, of higher fidelity modeling via the components that make up GPBC. GPGLM illuminates the value of bias correction, whereas GLM (which does not include a bias-correction) illustrates the value of surrogate modeling, as opposed to using raw simulations.  Climatological models, like OGP and GPBC without $\phi$, are often hard to beat on longer time scales \citep{thomas2020near}, i.e., in situations where dynamics are more climate- than weather-driven.  Weather forecasts become rapidly more inaccurate more than a week into the future.  

\subsection{Forecasting exercise}

We simulated one year of daily forecasting beginning June 11 2022 treating a corpus of model runs and sensor measurements from October 3 2020 through June 10 2022 as training, augmented daily.  GPBC and GPGLM model fitting is as described in Section \ref{sec:method}.  Training data for OGP consisted only of sensor measurements comprising the inital training period before June 10 2022. We present our results in several views.

\begin{figure}[ht!]
\includegraphics[scale=.31,trim=20 10 0 0]{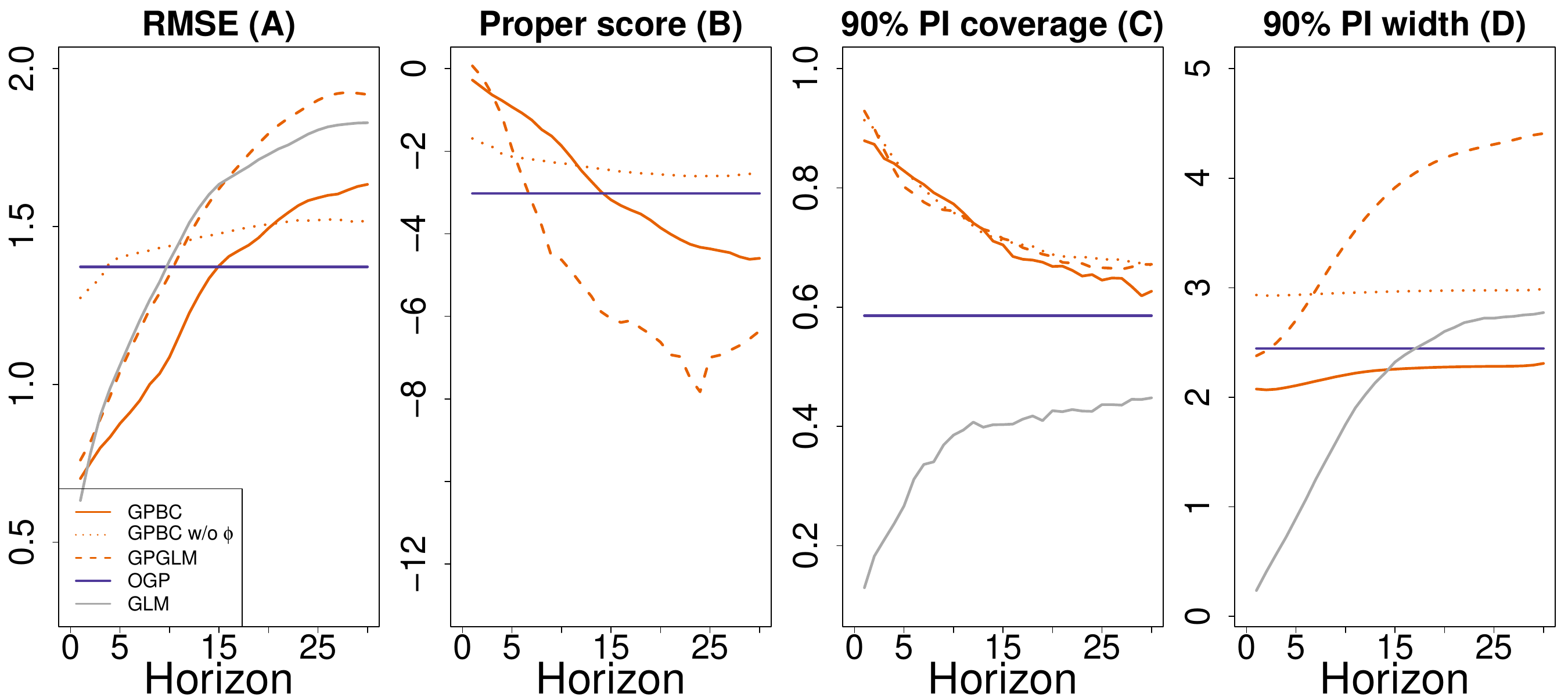}
\caption{RMSE ($^\circ$C), proper (log) score, and 90\% PI coverage, and PI interval width from left to right respectively, summarizing from out-of-sample forecasts for all competing models over horizon. Scores of GLM were omitted from the panel B because they were many orders of magnitude lower than other competitors. }
\label{fig:threePanelMetrics}
\end{figure} 

Figure \ref{fig:threePanelMetrics} shows RMSE, score, coverage, and PI width for each horizon, aggregating over all depths and days in the forecasting period.  First consider RMSE in panel (A), where lower is better.  Observe that every method outperforms OGP, the purely sensor-based climatological benchmark, at short horizons. This is a testament to the value of NOAA-GLM in the short term. However, OGP outperforms every other method after about two weeks out.  We suspect this is because NOAA forecasts are unreliable at long horizons.  In the shorter term, those methods that are most influenced by GLM forecasts perform best.  Standalone GLM forecasts or GPGLM (with the surrogate) have similar performance. Improvements in accuracy  due to the bias-correction in GPBC (with $\phi$) is clearly evident, being consistently better than both GPGLM and GLM. However, after two weeks climatological forecasts (OGP) win. Omitting $\phi$ compromises the influence of GLM forecasts, which results in some advantages as well as disadvantages.  Over short horizons, this comparator underwhelms, but at longer horizons its forecasts are more robust, similar to the purely climatological model OGP.  

Next, consider panel (B), which summarizes scores over the horizon (higher is better).  Overall trends here are similar to the RMSE panel, with some notable exceptions.  The raw GLM's score falls outside the lower range of the $y$-axis, so a line for this comparator does not appear on the plot.  Although it was reasonably accurate in terms of RMSE (left panel), it has poor UQ, which we shall discuss next with panels (D--E).  Again, our flagship GBPC comparator performs best for the first two weeks, except horizon one, where the surrogate-assisted GLM (GPGLM) has a better score than GPBC.  However, like raw GLM, this comparator is poor at higher horizons.  Observe that GPBC without wins for horizons longer than two weeks.  In particular, it beats the purely sensor-based climatological competitor OGP.

Panel (C) summarizes empirical coverage from 90\% predictive intervals over horizon.  Only the methods using surrogate-assisted NOAA-GLM forecasts can achieve nominal coverage (at 90\%) at any horizon.  Raw-GLM dramatically under-covers.  Its surrogate (GPGLM), by contrast, is able to correctly quantify uncertainty by accessing the entire corpus of simulations across the years. Although empirical coverage drops off from nominal at higher horizons, models that incorporate GLM forecasts (GPBC and GPGLM) always outperform OGP, the climatological benchmark. Additional insight into these coverage results is provided by panel (D).  Observe that the PIs of raw GLM forecasts are much narrower than those from other competitors, especially at early horizons. We also know (e.g.,~Figure \ref{fig:BCnoBCtrainperiodNOOGP}) that standalone GLM simulations and forecasts can exhibit pronounced bias. Taken together, this explains raw GLM's poor scores (panel B). Also note that PIs resulting from GPGLM are twice as wide as those of GLM for all horizons. Although GPBC and GPGLM have similar PI coverage, PIs of GPBC are much narrower than those of GPGLM. In fact, at horizon 30, PI width of GPBC (2.2) is nearly half that of GPGLM (4.2). 
Together, results show that bias-correction not only increases accuracy (via RMSE in panel A) but also results in narrower PIs. Note that although GPBC utilizes a homoskedastic GLM surrogate, PI interval width expands only slightly over horizon because the bias correction is homoskedastic.

Section D of the Supplementary Material \citep{HolthuijzenSupplement} provides additional analysis including stratification over depth, and a decomposition of accuracy and bias over day of year. 


\section{Conclusion}
\label{sec:discuss}

We studied forecasting lake temperatures 1 to 30 days into the future, bias-correcting and surrogate modeling a computer simulation campaign using a lake ecosystem model, GLM.  Our framework is thrifty in spite of massive data and cubic bottlenecks, enables iterative forecasting, and accounts for input-dependent variance.  When compared to other methods, without all of these ingredients, forecasts produced by our bias-corrected Gaussian process surrogate (GPBC) are typically more accurate and deliver superior uncertainty quantification (UQ), especially for forecasts up to two weeks ahead. However, climatological forecasts remain more accurate for predictions beyond two weeks, and in our future work we intend to explore possible reasons for this.
Further discussion and additional thoughts for future work are provided in Section E of the Supplementary Material \citep{HolthuijzenSupplement}.

\section{Data and code availability}
Code to replicate all modeling and figures is available in a Github repo: \url{https://github.com/maikeh7/Surrogate_Assisted_GLM}. All data  may be found in a Zenodo repo: \url{https://zenodo.org/uploads/10028017}.

\begin{acks} We thank the Virginia Tech Center for Ecosystem Forecasting team, who provided helpful feedback on this project. Members of the Center, Freya Olsson and Mary Lofton, edited earlier versions of this manuscript. We would also like to thank the Reservoir Group team who provided the lake temperature data analyzed in this study (data are available in the Environmental Data Initiative repository at https://doi.org/10.6073/pasta/7541e8d297850be7c613d116156735a9). 

\end{acks}

\begin{funding}
This project was supported by U.S. National Science Foundation grants DBI-1933016, EF-2318861, and DEB-2327030.  
\end{funding}

\begin{supplement}
\stitle{Supplementary material}
\sdescription{Additional figures, results, and discussion.}
\end{supplement}

\bibliographystyle{imsart-nameyear} 
\bibliography{refs}       

\end{document}